\documentclass[11pt, a4paper]{article}
\usepackage{jcappub}
\usepackage{subcaption}
\usepackage{verbatim}
\usepackage[numbers,sort&compress]{natbib}
\newcommand{\stochlss}{r^2(k)}

\newcommand{\fNL}{f_{\rm NL}}
\newcommand{\fNLeff}{\fNL^{\rm eff}}
\newcommand{\ftNL}{\tilde{f}_{\rm NL}}
\newcommand{\qftNL}{\left(\frac{q\ftNL}{1-q}\right)}

\newcommand{\Phix}{\Phi(\vec{x})}
\newcommand{\phix}{\phi_G(\vec{x})}
\newcommand{\psix}{\psi_G(\vec{x})}

\newcommand{\tpc}{(2\pi)^3}
\newcommand{\Mn}{\mathcal{M}_n}

\newcommand{\fkk}{\left( \frac{k}{k_0}\right)}

\newcommand{\MM}{\mathcal{M}_{3}}
\newcommand{\dmuint}{\int_{-1}^1 d\mu}

\newcommand{\Msunh}{h^{-1}M_{\odot}}
\newcommand{\Mpch} {\; \rm Mpc/h}
\newcommand{\hMpc} {\; h\rm Mpc^{-1}}
\newcommand{\klabel} {k\;(h\;\rm Mpc^{-1})}

\title{Higher moments of primordial non-Gaussianity and N-body simulations}

\author[a]{Saroj Adhikari,}
\author[a]{Sarah Shandera,}
\author[b]{Neal Dalal}

\affiliation[a]{Institute for Gravitation and the Cosmos, The Pennsylvania State University, University Park, PA 16802, USA}
\affiliation[b]{Department of Astronomy, University of Illinois, 1002 W. Green St., Urbana, IL 61801, USA}

\emailAdd{sza5154@psu.edu}
\emailAdd{shandera@gravity.psu.edu}
\emailAdd{dalaln@illinois.edu}

\abstract{We perform cosmological N-body simulations with non-Gaussian initial conditions generated from two independent fields. The dominant contribution to the perturbations comes from a purely Gaussian field, but we allow the second field to have local non-Gaussianity that need not be weak. This scenario allows us to adjust the relative importance of non-Gaussian contributions beyond the skewness, producing a scaling of the higher moments different from (and stronger than) the scaling in the usual single field local ansatz. We compare semi-analytic prescriptions for the non-Gaussian mass function, large scale halo bias, and stochastic bias against the simulation results. We discuss applications of this work to large scale structure measurements that can test a wider range of models for the primordial fluctuations than is usually explored.
}

\preprint{IGC-14/2-1}

\begin{document}
\maketitle

\section{Introduction}
Recently, the {\it Planck} satellite mission reported tight constraints on primordial non-Gaussianity from measurements of the Cosmic Microwave Background (CMB) \cite{PlanckCollaboration2013}. They find the amplitude of the local, equilateral and orthogonal type bispectrum to be $\fNL^{\rm local}=2.7\pm5.8$, $\fNL^{\rm equil}=-42\pm75$, and $\fNL^{\rm ortho}=-25\pm39$ respectively at $1\sigma$ level. While these results show that the primordial fluctuations were remarkably Gaussian, they still leave room for interesting signatures of primordial physics to be found in statistics beyond the power spectrum: minimal, single field models predict non-Gaussianity that is about two orders of magnitude below these constraints. Non-Gaussianity is such an informative tool for inflationary physics that it is crucial to push observational bounds as far as possible. A number of forecasts have shown that the use of clustering data from future large scale structure surveys can constrain $\fNL^{\rm local}$ with $\sigma(\fNL)\approx (1-10)$ \cite{Pillepich2012, Cunha2010, Oguri2009, Giannantonio2012,Merloni:2012uf, Amendola:2012ys}. In addition, the galaxy bispectrum is a promising way to probe many shapes to the $\fNL\sim\mathcal{O}(1-10)$ level \cite{Jeong:2009vd,Baldauf:2010vn,Tasinato:2013vna}. The constraints from large scale structure (LSS) ---whether consistent with the CMB measurements or in tension --- will provide interesting and useful complementary results. 

In the inflationary scenario, non-Gaussian signatures in the primordial density fluctuations depend on the details of interactions of the inflaton or other fields relevant for generating the perturbations; therefore, any detection (or the lack) of primordial non-Gaussianity tells us about the dynamics of those fields. Furthermore, if non-Gaussianity is detected, we expect to see patterns in the correlations that are consistent with perturbation theory (in contrast to apparently independent statistics at each order, for example). The simplest such pattern is in the relative amplitudes of higher order statistics, or, how the amplitudes of the correlation functions of the gravitational potential, $\langle\Phi^n(\vec{x})\rangle$, scale with order $n$. The relative scaling of higher moments falls into a fairly narrow range of behaviors in inflationary models \cite{Barnaby2012}. 

Thanks to the non-linearity of structure formation, the statistics of objects in the late universe contain information about the entire series of higher order moments of the initial density fluctuations. This is especially straightforward to see in number counts of galaxy clusters \cite{Loverde2007}. While information about higher moments is clearly non-trivial to extract when non-Gaussianity is weak, some constraints with current data are already affected by assumptions about higher order moments. For example, \cite{Shandera2013} reports constraints on primordial non-Gaussianity using a sample of 237 X-ray clusters from the ROSAT All-Sky Survey \cite{Trumper1993}. Their analysis clearly suggests that it is important to take the scaling of moments into account when deriving constraints on $\fNL$ from cluster number counts. The results presented in this paper will be useful for further studies in the same direction. Tighter constraints could likely be achieved with current data by combining the X-ray clusters with SZ detected clusters that have been separately used to constrain non-Gaussianity \cite{Benson2013, Williamson2011, Mana:2013qba}. In addition, the eROSITA survey \cite{Merloni:2012uf} and third generation surveys detecting clusters via the Sunyaev-Zel'dovich effect are expected to provide much larger samples of clusters in the next few years. It will be interesting to revisit the non-Gaussian analysis when that data becomes available. Other large scale structure statistics, including the power spectrum and bispectrum of galaxies, also contain contributions from higher order primordial statistics. Constraints on non-Gaussianity from those observables are still being developed, and will ultimately be very powerful (constraints obtained to date include \cite{Afshordi:2008ru, Slosar2010, Xia:2010pe, Xia:2011hj, Ross:2012sx, Karagiannis:2013xea, Giannantonio:2013uqa, Ho:2013lda, Agarwal:2013qta}). To make full use of them it is important to understand the signatures of the full spectrum of non-Gaussian models predicted from inflation.

In this paper, we will use a slight variant of the popular \textit{local} ansatz of primordial non-Gaussianity in order to generate two different scalings of the $n$-point functions and study how the scalings can be distinguished. For the usual local ansatz with \textit{hierarchical} scaling, many groups have already performed simulations \cite{Pillepich:2008ka, Grossi:2009an, Giannantonio:2009ak, Loverde2011, D'Amico:2010ta, Wagner2010} and found that the semi-analytic prescription of \cite{Loverde2007} for the non-Gaussian mass function works well. Here we will also test the validity of the mass function proposed in \cite{Barnaby2012} and used in \cite{Shandera2013} for another scaling that is well motivated by inflationary models. In addition to looking at the mass function, we will also look at a signature of the shape (momentum dependence) of local type non-Gaussianity using the scale dependent halo bias \cite{Dalal2008}, and at the stochastic bias on large scales \cite{Tseliakhovich2010}. Subleading contributions to the bias and leading contributions to the stochastic bias are sensitive to non-Gaussianity beyond the skewness. 

The rest of the paper is structured as follows: in Section \ref{sec:Model} we introduce our two field model and the scaling of the higher moments. In Section \ref{sec:Theory} we present the analytical calculations for the mass function of halos, bias, and stochasticity (with details relegated to the appendices). In Section \ref{sec:sim} we describe our simulations, and then compare the output from simulations to the analytical calculations in Section \ref{resultssection}. Finally, we conclude in Section \ref{sec:conclusion}.

\section{Model}
\label{sec:Model}
In the usual \textit{local} ansatz, the Bardeen potential $\Phi(x)$ has non-Gaussian statistics thanks to a contribution from a non-linear, local function of a Gaussian field $\psix$:
\begin{eqnarray}
 \Phi(\vec{x}) = \psi_G(\vec{x}) + \fNL \left( \psi_G(\vec{x})^2 - \langle
\psi_G(\vec{x})^2\rangle \right)\;.
\label{localmodel}
\end{eqnarray}
Here $\fNL$ parametrizes the size of the non-linear term and the level of non-Gaussianity. Non-Gaussianity of this type is usually thought of as produced by a light field that is not the inflaton \cite{Linde:1996gt, Moroi:2001ct, Lyth2001, Enqvist:2001zp, Dvali:2003em,Zaldarriaga:2003my}, and in fact cannot be generated by single field inflation proceeding along the attractor solution with modes in the Bunch-Davies vacuum \cite{Creminelli:2004yq, Pajer2013}.

The two-point statistics (the power spectrum) of the fluctuations are well measured from CMB observations for about three decades in scale \cite{Keisler:2011aw, Sievers:2013ica, Ade:2013zuv}. The amplitude of the three point function gives the skewness of the distribution of $\Phi$, and is also well-constrained by the CMB \cite{PlanckCollaboration2013}. Higher order correlation functions are more difficult to measure in the data. We would like to see if we can get some handle on the structure of the scaling of these higher order statistics of $\Phi$. For this purpose, we define the dimensionless moments $\Mn$:
\begin{eqnarray}
 \Mn = \frac{\langle \Phix^n \rangle_c}{\langle \Phix^2 \rangle_c^{n/2}}
 \label{eq:Mn}
\end{eqnarray}
where the subscript $c$ indicates that we take the connected part of the $n$-point function.
\paragraph{Two scalings:}
For the \textit{local} model given by Eq. (\ref{localmodel}), the moments scale approximately as
\begin{equation}
\label{eq:hierMn}
\mathcal{M}^{\rm hier}_n \approx A_n \left(\frac{\mathcal{M}_3}{6}\right)^{n-2}\;,\;\;\; n>2
\end{equation}
where $A_n =2^{n-3}n!$ comes out of combinatorics. The numerical coefficients in this scaling are not quite precise because of the difference in integrals over momenta at each order $n$ (see Appendix \ref{integrals}), but the {\it parametric} dependence on the amplitude of the skewness and the total power is fixed. This is the behavior of the moments (which we label `hierarchical' scaling) for $\fNL$ not too large i.e. when $2\fNL^2 \sigma^2 \ll 1$, where $\sigma^2=\langle \psix^2 \rangle$. However, if $2\fNL^2\sigma^2 \gg 1$, the moments scale differently:
\begin{equation}
\label{eq:feederMn}
\mathcal{M}^{\rm feeder}_n \approx B_n \left( \frac{\mathcal{M}_3}{8}\right)^{n/3}\;,\;\;\;n>2 
\end{equation}
where $B_n = 2^{n-1}(n-1)!$. In the single source case, this is far too non-Gaussian to be consistent with observations. Therefore, we consider the following two source model \cite{Lyth2006, Tseliakhovich2010, Barnaby2012}:
\begin{eqnarray}
 \Phix = \phix + \psi_G(\vec{x}) + \ftNL\left( \psi_G(\vec{x})^2 - \langle
\psi_G(\vec{x})^2\rangle\right)
\label{twofieldmodel}
\end{eqnarray}
where the two Gaussian fields $\psi_G(\vec{x})$ and $\phi_G(\vec{x})$ are uncorrelated i.e. $\langle \phi_G(\vec{x}) \psi_G(\vec{x})  \rangle=0$. 

To express the correlations in the gravitational potential $\Phi$ in terms of the amplitude of fluctuations in just one of the source fields we define
\begin{equation}
q = \frac{P_{\psi, G}(k)}{P_{\psi, G}(k)+P_{\phi}(k)}\;,
\label{eq:q}
\end{equation}
the ratio of the contribution of the {\it Gaussian} part of the field $\psi$ to the total Gaussian power. $P_{\psi,G}(k)$ is defined by
\begin{equation}
\langle \psi_G(\vec{k}) \psi_G(\vec{k}^{\prime})\rangle=(2\pi)^3\delta^3_D(\vec{k}+\vec{k}^{\prime})P_{\psi, G}(k)\;.
\end{equation}
For simplicity we assume both Gaussian components have constant, identical spectral indices but different amplitudes: $P_{\phi}(k)= 2\pi^2 \mathcal{P}_{\phi}(k)/k^3$ with $\mathcal{P}_{\phi}(k)=A_{\phi}\fkk^{\gamma}$, and similarly for $P_{\psi,G}$ (with the same index $\gamma$ but a different amplitude $A_{\psi}$). With this choice, $q$ is a scale-independent constant.

The power spectrum, bispectrum, and trispectrum in our model are:
\begin{eqnarray}
 P_{\Phi}(k) &=& \left[ \frac{1}{q}  + \ftNL^2 I_1(k) \mathcal{P}_{\psi,G}(k)\right] P_{\psi,G}(k)\nonumber \\
 &=& \left[\frac{1}{1-q} + \qftNL^2 I_1(k)\mathcal{P}_{\phi}(k)\right]P_{\phi}(k) \nonumber \\ \label{eq:powerspectrum} \\
 B(k_1,k_2,k_3) &=& 2 \ftNL \left[P_{\psi,G}(k_1) P_{\psi,G}(k_2) +
2\; {\rm perm} \right]\nonumber \\ & &+2 \ftNL^3 \left[ \int \frac{d^3\vec{p}}{\tpc}
P_{\psi,G}(p) P_{\psi,G}(|\vec{k_1}-\vec{p}|) P_{\psi,G}(|\vec{k_2}+\vec{p}|) +
3\;{\rm perm} \right] \label{eq:bispectrum} \\
T(k_1, k_2, k_3, k_4) &=& 2 \ftNL^2 \left[ P_{\psi,G}(k_1)
P_{\psi}(k_2) P_{\psi,G}(|\vec{k_1}+\vec{k_3}|) + 23 \; {\rm perm} \right] \nonumber \\
& &+  \ftNL^4 \left[ \int \frac{d^3\vec{p}}{\tpc} P_{\psi,G}(p)
P_{\psi,G}(|\vec{k_1}-\vec{p}|)
P_{\psi,G}(|\vec{k_2}+\vec{p}|)P_{\psi,G}(|\vec{k_2}+\vec{k_4}+\vec{p}|) 
\right.\nonumber \\ & &\left.+ 47 \; {\rm perm} \right] \label{eq:trispectrum}
\end{eqnarray}
where we have defined $I_1(k)$ as
\begin{eqnarray}
I_1(k) &=& 
\int_{\frac{k_{min}}{k}}^{\frac{k_{max}}{k}} du
\int_{-1}^1
d\mu \left[ u^{\gamma-1} (1 + u^2 + 2 \mu u )^{\frac{\gamma-3}{2}}\right]
\end{eqnarray}
Here, $k_{min}=2\pi/L$ is the infrared cutoff for a boxsize of $L$. In general, we do not know of the size of the universe beyond our observable universe, but for our purposes, the simulation box size $L$ is the natural choice. $k_{max}$ is the scale leaving the horizon at the initial epoch \cite{Lyth2007}. The above integrals converge for large values of $k_{max}$. For the computations to compare with simulations we set $k_{max}=N_p^{1/3} k_{min}$, where $N_p$ is the number of particles in a simulation. To arrive at the expressions quoted above, we have used
\begin{eqnarray}
\int \frac{d^3\vec{p}}{\tpc} P_{\psi}(p) P_{\psi}(|\vec{k}-\vec{p}|) &=& \frac{1}{2}I_1(k)\mathcal{P}_\psi(k)P_{\psi}(k)\;.
\end{eqnarray}

In Equations (\ref{eq:powerspectrum}), (\ref{eq:bispectrum}) and (\ref{eq:trispectrum}) we have included terms that are usually sub-dominant in the case of single field, weakly non-Gaussian local ansatz. In our model, these terms are important when the field $\psi$ is strongly non-Gaussian. To discuss the observational constraints on $(q, \ftNL)$, let's consider $\mathcal{P}_\Phi(k)\approx10^{-9}$, $I_1(k)\approx 10$ and $\fNL\lesssim \mathcal{O}(10)$.  Observational constraint from small non-Gaussianity can be satisfied by making $(q^2\ftNL)\lesssim \mathcal{O}(10)$ and $(q\ftNL)^3 \lesssim \mathcal{O}(10^{9})$, in which case the Gaussian contributions ($P_{\phi}(k)+P_{\psi,G}(k)$) dominate the total power spectrum in Eq.(\ref{eq:powerspectrum}) as well. Notice that the non-Gaussian contribution to the power can shift the spectral index slightly, so that when the $\psi$ field is strongly non-Gaussian the measured spectral index is close to, but not identical to, the spectral index of the Gaussian components\footnote{The new, integral terms have slightly different shapes than the usual terms. The difference, however, is small---approximately described by $\ln{kL}$ which has a weak dependence on $k$. These terms are also infrared divergent. For the purpose of comparing with the results from N-body simulations, the box size $L$ of the simulation provides a natural cutoff \cite{Loverde2011}. We will only look at quantities well enough inside the volume that the arbitrary size $L$ doesn't affect our results.}, $n_s-1\neq\gamma$. 

If $q \ll 1$ and $(q\ftNL)^3 \ll 10^{9}$, then one can generate the feeder scaling in Eq.(\ref{eq:feederMn}) without being inconsistent with the current observations of power spectrum and bounds on non-Gaussianity. The feeder scaling dominates when  the condition $q\ftNL^2\gg10^9$ is satisfied. This scaling, or a hybrid between hierarchical and feeder scaling, arises in particle physics scenarios where a second, non-Gaussian field couples to the inflaton and provides an extra source for the fluctuations \cite{Barnaby2012, Chen2010}. However, those scenarios differ from the model here because they most often generate bispectra not of the local type\footnote{One can also argue on statistical grounds that our observable universe is unlikely to have local non-Gaussianity of the type written in Eq.(\ref{twofieldmodel}) with large $\tilde{f}_{\rm NL}$ if inflation lasts much longer than 55 e-folds \cite{Nelson2013, Loverde2013}).}. Here we are using the two field, local model primarily as a phenomenological tool, easy to implement in N-body simulations, to generate the feeder-type scaling of moments rather than as the output of a particular inflation model. The mass function is sensitive only to the integrated moments, not the shape, so this is a useful test of how different scalings affect the number of objects in a non-Gaussian cosmology.

\section{Abundance and Clustering Statistics}
\label{sec:Theory}
In this section we present the analytic predictions for the effect of locally non-Gaussian, two source initial conditions on the abundance and clustering of dark matter halos. 

At large scales, the evolution of the density contrast is well described by linear perturbation theory, and the density contrast in Fourier space at redshift $z$ is given by $\delta(\vec{k},z) = \alpha(k,z) \Phi(\vec{k})$, where
\begin{eqnarray}
 \alpha(k,z) = \frac{2}{3} \frac{k^2 T(k) D(z)}{H_0^2 \Omega_m}\;.
\end{eqnarray}
Here $T(k)$ is the transfer function, $D(z)$ is the growth function, $H_0$ is the Hubble scale today, and $\Omega_m$ is the energy density in matter today (compared to critical density). The smoothed density contrast, similarly, is $\delta_R(\vec{k},z) = W_R(k) \alpha(k,z) \Phi(\vec{k})$, where
\begin{eqnarray}
 W_R(k) = \frac{3 \sin({kR}) - 3 (kR) \cos({kR})}{(kR)^3}
\end{eqnarray}
is the smoothing function (here the Fourier transform of the real space top-hat). Note that we will generally suppress the $z$ dependence in $\alpha(k,z)$ and $\delta(\vec{k},z)$ in this paper, and usually write $\alpha(k)$ and $\delta(\vec{k})$ only. We can now compute the connected n-point functions of the smoothed density contrast in real space:
\begin{eqnarray}
\label{eq:deltaRn}
\langle \delta_R(\vec{x}) ^n\rangle_c(z) = \int \frac{d^3
\vec{k_1}}{\tpc}\dots\int \frac{d^3\vec{k_n}}{\tpc} \langle
\delta_R(\vec{k_1},z)\dots\delta_R(\vec{k_n},z)\rangle_c
\end{eqnarray}
and therefore the dimensionless moments of smoothed density fields: $\mathcal{M}_{n,R}=\frac{\langle\delta_R(\vec{x})^n\rangle_c}{\langle\delta_R(\vec{x})^2\rangle^{n/2}}$. Note that the dimensionless moments are redshift independent. Eq.(\ref{eq:deltaRn}) can be separated into two components: (i) of $\mathcal{O}(\ftNL^{n-2})$ and (ii) of $\mathcal{O}(\ftNL^n)$, corresponding to contributions to the hierarchical and feeder scalings respectively. Some of these integrals, with a brief summary of our method to evaluate them, are given in Appendix \ref{integrals}.

\subsection{Mass function}
We follow previous studies of non-Gaussian mass functions in that we calculate the ratio $R=\frac{n_{NG}(M,z)}{n_G(M,z)}$ of the number density in the presence of non-Gaussianity, $n_{NG}$, to the number density for Gaussian initial conditions, $n_{G}$, for a particular halo mass $M$ at some redshift $z$ using the Press-Schechter formalism \cite{Press1974}. The fractional volume $F(M)$ of dark matter in the collapsed structures (halos) is proportional to $\int_{\delta_c}^{\infty} P(\delta_M) d\delta_M$, where $\delta_c$ is the critical value of the smoothed density contrast $\delta_M$ above which the dark matter in a region collapses to form halos and $P(\delta_M)$ is a probability density function (PDF). Here we have written the smoothing scale in terms of the mass $M$ rather than the smoothing radius $R$; they are simply related by $M=\frac{4}{3}\pi R^3 \rho_m$, where $\rho_m$ is the mean matter density of the universe. Then, the number density (mass function) is given by:
\begin{eqnarray}
\frac{dn}{dM} = \frac{dF(M)}{dM} \times \frac{\rho_m}{M}
\label{massfneqn}
\end{eqnarray}
For a Gaussian PDF, one can easily perform the integration to find a prediction for the mass function. However, the result is known to be only an approximation and in practice Gaussian mass functions are calibrated on simulations \cite{Tinker2008}. 

To apply the method above in case of non-Gaussian initial conditions we need a non-Gaussian PDF. The Petrov expansion \cite{petrovpaper} (which generalizes the Edgeworth expansion \cite{Blinnikov:1997jq, Bernardeau:2001qr}) expresses a non-Gaussian PDF as a series in the cumulants of the distribution. In terms of $\mathcal{M}_{n,R}$'s, this is:
\begin{eqnarray}
 P(\nu, R) &=& \frac{e^{-\nu^2/2}}{\sqrt{2\pi}} \left[ 1+ \sum_{s=1}^{\infty}
\sum_{k_m} H_{s+2r}(\nu) \prod_{m=1}^s \frac{1}{k_m!} \left(
\frac{\mathcal{M}_{m+2, R}}{(m+2)!}\right)^{k_m}\right]
\label{eq:pdf}
\end{eqnarray}
where $\nu=\frac{\delta_M}{\sigma_M}$, $\sigma_M=\sqrt{\langle\delta_M^2\rangle}$ and $H_n$'s are the Hermite polynomials defined as $H_n(\nu)=(-1)^n e^{\nu^2/2} \frac{d^n}{d\nu^n} e^{-\nu^2/2}$.  The second sum is over the non-negative integer members of the set $\{k_m\}$ that satisfy
\begin{equation}
\label{eq:Diophantine}
k_1+2k_2+\dots+sk_s=s\;.
\end{equation}
For each set $r\equiv k_1+k_2+\dots +k_s$. This series can be integrated term by term to obtain $F(M)$, and with Eq.(\ref{massfneqn}) gives ratio of non-Gaussian mass function to Gaussian mass function \cite{Shandera2013}
\begin{eqnarray}
 \frac{n_{NG}}{n_G} \approx 1+ \frac{F_1^{h,f'}(M)}{F_0'(M)} + 
\frac{F_2^{h,f'}(M)}{F_0'(M)}+\dots
\label{nng}
\end{eqnarray}
where 
\begin{eqnarray}
 F_0'(M) = -\frac{\nu_c'(M)}{\sqrt{2\pi}} e^{-\frac{1}{2} \nu_c(M)^2}\;, \;\;\;\nu_c=\frac{\delta_c}{\sigma_M}.
\end{eqnarray}
The two superscripts $h,f$ indicate that the set of terms that are of the same order depends on whether higher order cumulants have hierarchical ($h$) or feeder ($f$) scaling. Formally grouping terms assuming the scalings in Eq(\ref{eq:hierMn}) or Eq.(\ref{eq:feederMn}) are exact gives (for $s\geq1$)
\begin{eqnarray}
F_s^{h'}(\nu) &=& F_0' \sum_{\{k_m\}_h}\!\!\left\{\!H_{\small{s+2r}}\!\prod_{m=1}^s 
\frac{1}{k_m!} \left(\frac{\mathcal{M}_{m+2,R}}{(m+2)!}\right)^{k_m}\!\!\!+H_{s+2r-1}\frac{\sigma}{\nu}\frac{d}{d\sigma}\left[\prod_{m=1}^s\frac{1}{k_m!}\left(\frac{\mathcal{M}_{m+2,R}}{(m+2)!}\right)^{k_m}\right] \right\}\nonumber \\ \\
F_s^{f'}(\nu) &=& F_0' \sum_{\{k_m\}_f}\!\!\left\{H_{s+2}\prod_{m=1}^s\frac{1}{k_m!} \left(\frac{\mathcal{M}_{m+2,R}}{(m+2)!}\right)^{k_m}\!\!\!+H_{s+1}\frac{\sigma}{\nu}\frac{d}{d\sigma}\left[\prod_{m=1}^s\frac{1}{k_m!}\left(\frac{\mathcal{M}_{m+2,R}}{(m+2)!}\right)^{k_m}\right] \right\}\;. \nonumber \\
\end{eqnarray}
The prime stands for the derivative with respect to the halo mass $M$. In the hierarchical case, the sets $\{k_m\}_h$ still satisfy Eq.(\ref{eq:Diophantine}), while for feeder scaling the $\{k_m\}_f$ are the sets of non-negative integer solutions to \mbox{$3k_1+4k_2+\dots+(s+2)k_s=s+2$}.

Eq.(\ref{nng}) assumes that the two point statistics of the smoothed linear density contrast (i.e $\langle \delta^2_M\rangle$) are the same for the non-Gaussian and the Gaussian cases (not that the {\it Gaussian} contributions to $\langle \delta^2_M\rangle$ are the same). However, this is difficult to maintain at all scales in our simulations. We require the two point clustering statistics to match at a particular scale $R=8 \Mpch$, but do not correct for the shift to the spectral index coming from the non-Gaussian term in the power spectrum. As a result, on scales other than $R=8 \Mpch$ we need to make a distinction between $F_{0,NG}'(M)$ for the non-Gaussian cosmology and $F_{0, G}'(M)$ for the Gaussian cosmology. In that case, there is an extra (mass or scale dependent) factor $f_1(M)$ in the ratio $\frac{n_{NG}}{n_G}$, where
\begin{eqnarray}
 f_1(M) &=& \frac{F_{0, NG}'(M)}{F_{0,G}'(M)} = \frac{\nu_{c, NG}'(M)}{\nu_{c,G}'(M)} e^{-\frac{1}{2}\left(\nu_{c,NG}^2-\nu_{c,G}^2\right)}
 \label{eq:f1M}
\end{eqnarray}
This factor is typically quite close to one. For example, in the mass range ($4\times10^{13} < M < 2\times 10^{15}$)$h^{-1}M_{\odot}$ at $z=1$, $0.995\lesssim f_1(M) \lesssim  1.001$ for single field $\fNL=500$ case. For our smallest feeder scaling simulation (with $\mathcal{M}_3\approx0.020$), in this mass scale and redshift range, $0.92 \lesssim f_1(M) \lesssim 1.02$. The factor deviates away from unity more at larger redshifts and at mass scales far from $M\approx 1.61\times 10^{14} h^{-1} M_{\odot}$.

In addition, the derivation above assumes the same constant of proportionality between $F(M)$ and $\int_{\delta_c}^{\infty} P(\delta_M) d\delta_M$ regardless of the level of non-Gaussianity. The standard Press-Schechter constant of proportionality is two, but for the non-Gaussian case it is reasonable to fix the constant by requiring $\bar{\rho} = \int_0^\infty M \frac{dn}{dM} dM$. Gaussian and non-Gaussian cosmologies with identical $\sigma_8$ will have slightly different normalization factors. This factor shifts further away from 2 as the level of non-Gaussianity in the initial conditions is increased. Integrating various truncations of the expanded PDFs indicates we expect a difference from 2 between about 0.5\% and 2\% for the amplitudes of non-Gaussianity we consider in this work. So, we will introduce an extra factor $f_2$ that multiplies our analytical mass function to fit with the simulation results.

\subsection{Large scale bias}
On large scales, where density fluctuations are in the linear regime, the clustering of halos is expected to trace the clustering of the underlying matter field. The proportionality constant relating halo clustering to matter clustering is called the halo bias. Local type non-Gaussianity can modify halo bias, compared to Gaussian universes, by coupling the amplitude of short wavelength modes to that of long wavelength modes. Since the coupling occurs in the gravitational potential field (with constant amplitude), rather than in the density field, local type non-Gaussianity introduces a new, scale-dependent term relating the power spectrum of halos to the power spectrum of the linear dark matter field. On large scales (small wave numbers) the non-Gaussian term can dominate and the analytic prediction for the bias is relatively simple.

The potential use of the halo (or galaxy) bias as a probe of primordial non-Gaussianity was first demonstrated in \cite{Dalal2008}. An analytic derivation capturing the first order effect of non-Gaussian initial conditions had been presented much earlier in \cite{Matarrese:1986et} and clarified and improved following the Dalal et al result in \cite{Matarrese:2008nc, Desjacques2011, Desjacques:2011mq}. The halo bias in models with two sources for the primordial fluctuations was considered in \cite{Tseliakhovich2010, Baumann2012}, and in  \cite{Yokoyama2011} which gives some theoretical predictions for the model we consider here. In addition, \cite{Smith2011} previously performed N-body simulations for a model where the kurtosis was larger than in the single field case by a factor of two. That work corresponds to our scenario with $q=0.5$. Here we are primarily interested in values of $q$ that are very small so that the scaling is feeder type, but we also consider cases with $q\approx0.1$ which have intermediate scaling. Measurements of the power spectra of several different galaxy populations have been used to place constraints on primordial non-Gaussianity, at roughly the $|\sigma(f_{\rm NL})|\sim\mathcal{O}(25-200)$ level, depending on the population and treatment of systematic errors \cite{Afshordi:2008ru, Slosar2010, Xia:2010pe, Xia:2011hj, Ross:2012sx, Karagiannis:2013xea, Giannantonio:2013uqa, Ho:2013lda, Agarwal:2013qta}.

The general form of the large scale bias for our two-field model Eq.(\ref{twofieldmodel}) can be calculated using the peak-background split formalism \cite{Slosar2010, Smith2011}. Appendix \ref{app:bias} has the peak-background-split calculation that results in Eq.(\ref{eq:hmbias}) as the expression of bias. In case of small, local non-Gaussianity ($q\ll1$ for the case with a second, strongly non-Gaussian field), the expression for large scale bias reduces to:
\begin{eqnarray}
 P_{hm}(k) &=& \left[b_{\rm all\; sources} + \frac{2\delta_c(b_{\rm NG\; source}-1)}{\alpha(k)} \left( q^2\ftNL +(q\ftNL)^3
\mathcal{P}_{\Phi}(k) I_1(k) \right)\right] P_{mm}(k)\;. \nonumber \label{eq:biassimple} \\
\end{eqnarray}
Here we have used subscripts on the bias coefficients to emphasize that the scale-independent term depends on all sources of the fluctuations ($b_{\rm all\; sources}$), while the scale-dependent term depends only on those sources with a primordial non-Gaussian component ($b_{\rm NG\; source}$). Recall that in the more frequently quoted expression with a single, non-Gaussian source, these two bias coefficients are both equal to the Gaussian bias plus scale-independent corrections proportional to the level of non-Gaussianity. When multiple sources are present, the first bias coefficient can be split into terms which are include the Gaussian bias for each source, while the bias coefficient in the second, scale-dependent term is to lowest order the Gaussian bias for the {\it non-Gaussian source}. In addition, Eq.(\ref{eq:biassimple}) demonstrates that the large scale halo bias is a probe of $q^2\ftNL$ for the hierarchical scaling and a probe of $(q \ftNL)^3$ for the feeder scaling. In other words, the dominant contribution to the non-Gaussian bias is proportional to the amplitude of the bispectrum as expected.  

To compare the analytic expressions against simulation results, we will use the size of the simulation volume to truncate integrals in the infrared and will not fit the power spectrum on scales very close to the simulation box size ($k \lesssim0.007 \hMpc$)\footnote{First, for $k \lesssim 0.007 \hMpc$ i.e modes approaching the scale of the box size of our simulations, the sample variance is large. Second, the bias for feeder scaling depends on $I_1(k)$ which has a sharply decreasing behavior near $k_{min}$ of the simulation; this effect runs with the box size. If we are interested in scales near the $k_{min}$ ($k \lesssim 0.007 \hMpc$) for our feeder scalings, then we will either have to do simulations with larger box sizes, or generate initial conditions differently such that when averaged over many realizations of the initial conditions we don't get this feature.}. 

\subsection{Large scale stochasticity}
Using the peak-background-split method, we can also calculate the halo-halo power spectrum $P_{hh}(k)$ in terms of the underlying matter distribution $P_{mm}(k)$. We can then define the large scale stochasticity, or cross-correlation coefficient as
\begin{eqnarray}
 \stochlss &=& \frac{P_{hm}^2(k)}{P_{hh}(k) P_{mm}(k)}
 \label{eq:stochasticity}
\end{eqnarray}

The calculations and the corresponding expressions for $P_{hm}(k)$ and $P_{hh}(k)$ are given in Appendix \ref{app:bias}. Peak-background split calculations suggest that in the large scale limit (small $k$), $\stochlss$ gives the fraction of power in the initial conditions from contributions of the field with non-Gaussianity (i.e. $1-(P_{\phi}(k)/P_{\Phi}(k))$. For small non-Gaussianity, in case of hierarchical scaling, this is $\approx q$, while in case of feeder scaling, this is $\approx (q \ftNL)^2 \mathcal{P}_{\phi}(k)$. For the single field cases, there is no large scale stochasticity ($\stochlss=1$). We will test these expectations against results from simulations. Large scale stochasticity for the Gaussian case and non-Gaussian case with hierarchical scaling was discussed in \cite{Smith2011}; their comparison with simulations showed a small discrepancy between their analytic expression and the simulation results. Our results below indicate one possible source for at least some of that discrepancy.

\section {Simulations}
\label{sec:sim}
The simulations for this project were performed using the popular \texttt{GADGET-2} code \cite{gadget2}. The initial conditions were generated using second order Lagrangian perturbation theory (\texttt{2LPT}) \cite{Crocce:2006ve}. We used the code from \cite{Crocce:2006ve} for generating local type (single field) non-Gaussian initial conditions using \texttt{2LPT} and modified the code---discussed next---to generate initial conditions for our two field model. 
\paragraph{Initial conditions:} First, two Gaussian random fields ($\phix$ and $\psix$) were generated using the power spectrum of our fiducial Gaussian cosmology ($n_s=0.96, \sigma_8=0.8, \Omega_m=0.27, \Omega_\Lambda=0.73$). The amplitude of the power spectrum for the $\psix$ and $\phix$ fields were multiplied by the appropriate factors $q$ and $(1-q)$ respectively. Then, the $\psix$ field was squared and multiplied by $\ftNL$. The total non-Gaussian field $\Phix$ was obtained by adding the three components as in Eq. (\ref{twofieldmodel}). We ensured $\sigma_8$ of the generated non-Gaussian field was that of our specified cosmology for all of our parameter sets by renormalizing the $\Phix$ field by a factor of $\left(\sigma_8/\sigma_{8, \Phi}\right)^{0.5}$. These are the only modifications done to the \texttt{2LPT} code. The rest of the code generated the required displacement and velocity fields as usual. The redshift of all our initial conditions is $z_{\rm start}=49$.
\paragraph{N-body simulations:} The public version of the \texttt{GADGET-2} code was used to perform cosmological collisionless dark matter only simulations. All simulations were done with $(1024)^3$ particles in a cube of side $2400 h^{-1}Mpc$. This gives the mass of a single particle to be $9.65\times10^{11} h^{-1}M_{\odot}$. The force softening length was set to be $5\%$ of the inter-particle distance. Simulations with Gaussian initial conditions ($q=0$ and $\ftNL=0$) were also performed with the same seeds as the $\phi$ field (as it has the dominant contribution for small $q$) to compare with our feeder models; another set of Gaussian simulations was performed with $q=1$ and $\ftNL=0$ to compare with the hierarchical simulations. For each set of parameters listed in Table \ref{table:parameterspace}, we ran four simulations with different seeds. All simulation results reported in this paper are average over the four simulations. Similarly, the errors reported are the 1$\sigma$ standard deviation of the four simulations. The \texttt{AHF} halo finder \cite{ahf} was used to identify halos which were then used to get the mass function of dark matter halos and power spectra of halos (halo-matter cross power spectrum and halo-halo autospectrum). In all our analyses, we only use halos with number of dark matter particles $N_p\geq50$.

\paragraph{Parameter space of simulations:} Since our method of generating the feeder scaling produces a slightly different bispectrum shape than the hierarchical case, the scale dependence of $\MM$ for the two scalings also differ. For comparison purposes, we will define $\fNLeff$ at the scale of $R = 8 \Mpch$ corresponding to a halo mass of $1.61\times10^{14} h^{-1}M_{\odot}$, as the ratio:
\begin{eqnarray}
 \fNLeff = \frac{
\mathcal{M}_{3}(q,\ftNL)}{\mathcal{M}_{3}({q=1,\ftNL=1})}\bigg\rvert_{M=1.61\times 10^{14} h^{-1} M_{\odot}}
\label{fNL8}
\end{eqnarray}

In Table \ref{table:parameterspace}, we list the parameter sets of $(q, \ftNL)$ that we have simulated. Notice that these parameter sets are not consistent with the small $f_{\rm NL}^{\rm local}$ reported by the Planck mission from bispectrum measurements. However, it is necessary to use parameter sets with larger values of $\fNLeff$ in order to get useful results from N-body simulations. 

\begin{table}
\begin{center}
\begin{tabular}{|c|c|c|c|c|c|}
\hline
 Name & $\fNL^{\rm eff}$
(Eq:(\ref{fNL8})) & $q$ & $\ftNL$& $\mathcal{M}_3$ & $\mathcal{M}_{3,f}/\mathcal{M}_{3,h}$  \\ \hline
 \texttt{M993}& 993 & 0.1 & 50000 &0.290 & 2.144 \\ \hline
 \texttt{F677}& 677 & 0.00005 & $10^8$ & 0.198 & 4287  \\ \hline
 \texttt{H500}& 500 & 1 & 500 & 0.145 &0.0027  \\ \hline
 \texttt{M384}& 384 & 0.11925 & 20620 &  0.112& 0.5089  \\ \hline
 \texttt{F215}& 215 & 0.00003 & $10^8$ & 0.063 & 2923  \\\hline
 \texttt{F122}& 122 & 0.00003 & $8\times10^7$ & 0.036 &1933 \\ \hline
 \texttt{H99}& 99 & 1 & 99 & 0.029 & 0.0001  \\ \hline
 \texttt{F70} & 70 & 0.00003 & $6.5\times 10^7$ & 0.020 & 1303 \\ \hline
\end{tabular}
 \caption{Parameter space of our simulations. In the first column, we name our models following a simple naming convention. The first letter of the name stands for the type of scaling of the model: \texttt{F} stands for feeder scaling, \texttt{H} stands for hierarchical scaling, \texttt{M} stands for mixed scaling (when neither component is negligible). The number following  is the approximate $\fNLeff$, also listed in the second column. For example: \texttt{F70} means that the scaling of the model is feeder and has $\fNLeff=70$. The quantity $q$ in the third column is defined in Eq.(\ref{eq:q}), and gives the ratio of power in the Gaussian part of the $\psi$ contribution ($P_{\psi,G}$), to the total Gaussian power ($P_{\psi,G}+P_{\phi}$). The second last column is the dimensionless skewness $\MM$ computed at a halo mass scale $M=1.61 \times 10^{14} \Msunh$. The last column $\mathcal{M}_{3,f}/\mathcal{M}_{3,h}$ is the ratio of the dimensionless skewness $\MM$ from the feeder contribution to that of the hierarchical contribution and indicates the relative importance of the feeder term. }
\label{table:parameterspace}
\end{center}
\end{table}

\section{Results and Discussion}
\label{resultssection}
\paragraph{Mass Functions:} 
The hierarchical scaling has been considered a number of times already with the prediction from the Edgeworth-series formalism \cite{Loverde2007} providing good fit to the outputs from simulations \cite{Pillepich:2008ka, Grossi:2009an, Giannantonio:2009ak, Loverde2011, D'Amico:2010ta, Wagner2010}. Our focus will be on the feeder type scaling. 

In the following sections, we will use the mass function truncated up to $\mathcal{M}_5$. From error analysis (Appendix \ref{app:erroranalysis}), we see that we gain little by adding higher terms for the feeder mass function. To this order, for the feeder case, the ratio of non-Gaussian to Gaussian mass function is:
\begin{eqnarray}
 \left(\frac{n_{NG}}{n_G}\right)_{\rm feed} &=& f_1(M)  f_2^{\rm feed} \left[ 1 + \frac{\MM}{3!}H_3(\nu_c)- 
\frac{\MM'}{3!\nu_c'}H_2(\nu_c) +\frac{\mathcal{M}_{4}}{4!} H_4(\nu_c) \right. \nonumber \\ 
&& \left. - 
\frac{\mathcal{M}_{4}'}{4! \nu_c'} H_3(\nu_c) +\frac{\mathcal{M}_{5}}{5!} H_5(\nu_c) - 
\frac{\mathcal{M}_{5}'}{5! \nu_c'} H_4(\nu_c) \right]
\label{eq:feedmassfn}
\end{eqnarray}
where we have chosen to rewrite all the quantities in terms of the halo mass $M$ rather than $\sigma$ and $R$; the $M$ dependence of the moments $\mathcal{M}_n$'s and $\nu_c$ has been suppressed for clarity, as is done throughout the paper. We use this expression to fit to the results from N-body simulations.

Similarly, for hierarchical scaling of moments, the expression for the non-Gaussian mass function including terms upto $\mathcal{O}(\mathcal{M}_5)$ is:
\begin{eqnarray}
 \left(\frac{n_{NG}}{n_G}\right)_{\rm hier} &=& f_1(M) f_2^{\rm hier} \left[ 1 + \left( \frac{\mathcal{M}_{3}
H_3(\nu_{c})}{3!} - \frac{\mathcal{M}_{3}' H_2(\nu_{c})}{3! \nu_{c}'}\right) + \right. \nonumber \\
& & \left(\frac{
\mathcal{M}_{4} H_4(\nu_{c})}{4!} + \frac{\mathcal{M}_{3}^2
H_6(\nu_{c})}{2\times3!\times3!} \nonumber - \frac{
\mathcal{M}_{4}' H_3(\nu_{c})}{4!\nu_{c}'} - \frac{\mathcal{M}_{3}
\mathcal{M}_{3}' H_5(\nu_{c})}{3!\times3! \nu_{c}'} \right) \nonumber \\
&+& \left. \left( \frac{\mathcal{M}_{5} H_5(\nu_{c})}{5!} - \frac{\mathcal{M}_{5}' H_4(\nu_{c})}{5! \nu_{c}'}+ \frac{\mathcal{M}_{3} \mathcal{M}_{4} H_7(\nu_{c})}{4!3!} + \left(\frac{\mathcal{M}_{3}}{3!}\right)^3 \frac{H_9(\nu_{c})}{3!} \right. \right. \nonumber \\
 & & \left.\left. - \frac{(\mathcal{M}_4\mathcal{M}_3'+\mathcal{M}_3\mathcal{M}_4') H_6(\nu_c)}{3! 4! \nu_c'} - \frac{1}{3!} \left(\frac{\mathcal{M}_3}{3!}\right)^2 \frac{\mathcal{M}_3' H_8(\nu_c)}{3!\nu_c'} \right) \right]
\label{eq:hiermassfn}
\end{eqnarray}
where the $M$ dependence of $\mathcal{M}_n$ and $\nu_c$ has been suppressed for clarity. In the above mass function formulae for feeder scaling Eq.(\ref{eq:feedmassfn}) and for hierarchical scaling Eq.(\ref{eq:hiermassfn}), the expression for $f_1(M)$ is calculated for a given model using Eq.(\ref{eq:f1M}) and the $f_2$ factors are fit for each of our simulation results. Both of these factors approach unity for small non-Gaussianity. To compute the cumulants necessary to calculate the dimensionless moments $\mathcal{M}_n$ in the above formulae, we have used the Monte-Carlo method described in \cite{Loverde2011}. See Appendix \ref{integrals} for details.

Let us now compare and discuss the results from simulations and calculations from the Edgeworth series formulation. First, as a check of our simulations, we did a purely hierarchical scaling parameter set---single field, $\fNL=500$ simulation---which can be compared directly with previous works. Then, for the two source case, we study the parameter sets listed in Table \ref{table:parameterspace} which allow feeder scaling as well as mixed scaling (i.e, neither term in Eqs.(\ref{eq:bispectrum}),(\ref{eq:trispectrum}) is negligible). In this section, we will present the simulation results and the corresponding Edgeworth mass functions. We analyzed our mass function results with a simple two parameter ($\delta_c, f_2$) chisquare minimization procedure. The errors reported in the best fit values increase the reduced chisquare of the fit by unity when added to the best fit values.  We find that all simulations (\texttt{H99}, \texttt{H500}, \texttt{F70}, \texttt{F122}, \texttt{F215}) prefer a reduced $\delta_c\approx 1.4 - 1.5$ and different values for $f_2$, which are of the expected size and increase appropriately with $\mathcal{M}_3$. We will discuss the dependence of this factor on $\mathcal{M}_3$ and the type of scaling later. 

\begin{figure}
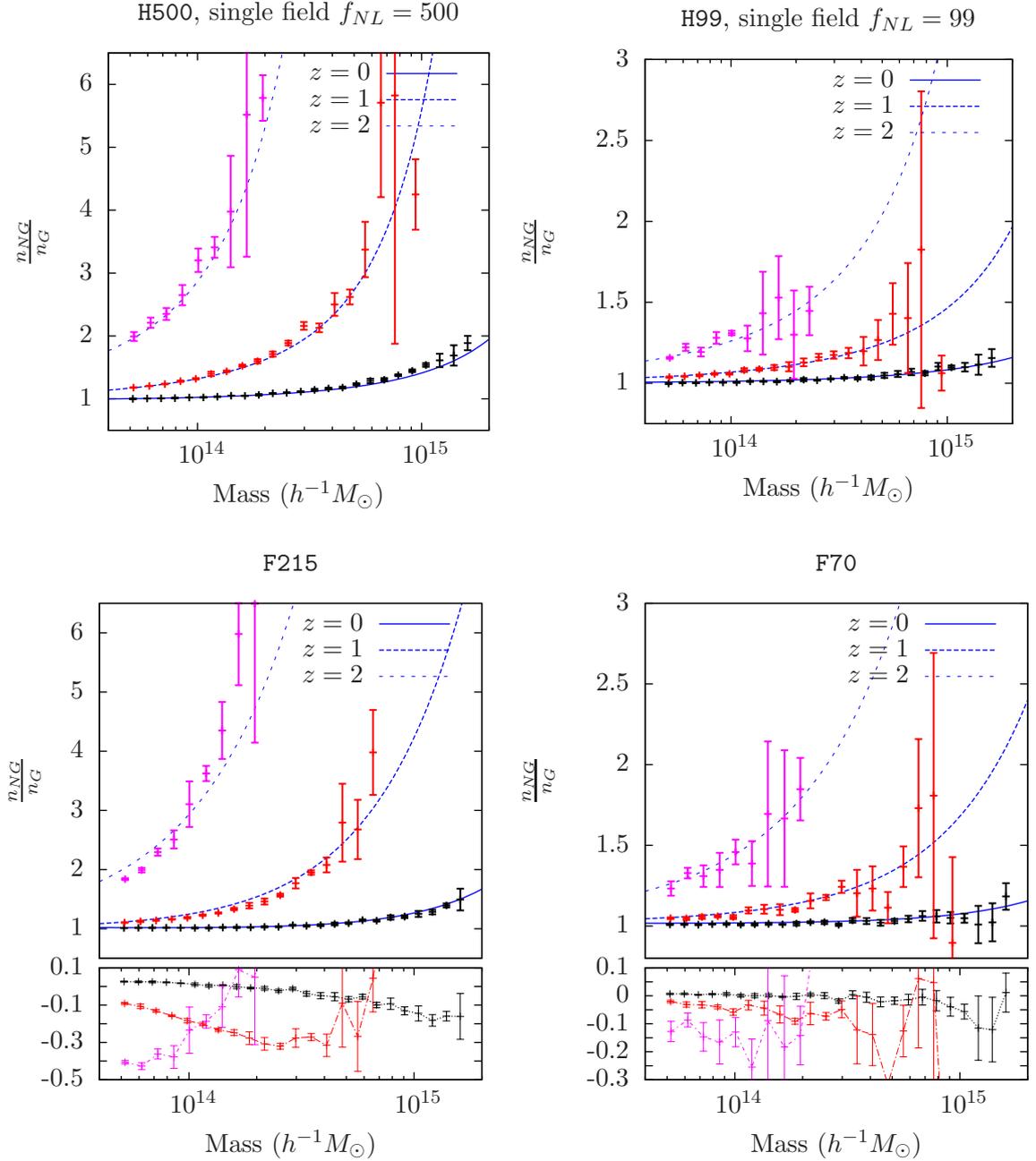

 \centering
 \begin{subfigure}[b]{0.48\textwidth}
  \centering
  \input{figures/fnl500.tex}
 \end{subfigure}
 \begin{subfigure}[b]{0.48\textwidth}
  \centering
  \input{figures/fnl99.tex}
 \end{subfigure}
\begin{subfigure}[b]{0.51\textwidth}
  \centering
  \input{figures/rccc.tex}
 \end{subfigure}
 \begin{subfigure}[b]{0.47\textwidth}
  \centering
  \input{figures/rsss.tex}
 \end{subfigure}
 \caption{\textit{Top}: the simulation results and semi-analytic prediction Eq.(\ref{eq:hiermassfn}) for hierarchical simulations. Left panel: $\mathcal{M}_3=0.145$ ($\fNL=500$) with $f_2=1.042$. Right panel: $\mathcal{M}_3=0.029$ ($\fNL=99$) with $f_2=1.009$. \textit{Bottom}: The simulation results and semi-analytic prediction Eq.(\ref{eq:feedmassfn}) for  feeder simulations. Left panel: $\mathcal{M}_3=0.063$ ($f_{\rm NL}^{\rm eff}=215$) and $f_2=1.043$. Right panel: $\mathcal{M}_3=0.020$  ($f_{\rm NL}^{\rm eff}=70$) with $f_2=1.012$. In addition, we also plot the fractional difference between the simulation results and the prediction from the truncated mass function Eq.(\ref{eq:feedmassfn}) but using ($\delta_c=1.686, f_2=1$), and assuming that $\mathcal{M}_n$ scale as Eq.(\ref{eq:feederMn}). A negative value means that the analytic mass function overpredicts the simulation result.}
 \label{mfnresults}
\end{figure}

First, let us present our mass function fits. We use $\delta_c=1.46$ obtained by performing chisquare minimization together for \texttt{H99, H500, F70, F122, F215} by forcing the same $\delta_c$ but allowing overall rescaling factors ($f_2$) for each case. This is a reasonable procedure if one interprets shifting $\delta_c$ as allowing departures from the assumption of spherical collapse, which should be relatively independent of the level of non-Gaussianity. However, the normalization requirement $\bar{\rho} = \int_0^\infty M \frac{dn}{dM} dM$ (which has not been analytically enforced) suggests that $f_2$ should depend on the level of non-Gaussianity. The top panel of Figure \ref{mfnresults} has our mass function results for two simulations with hierarchical scalings.  

\begin{figure}
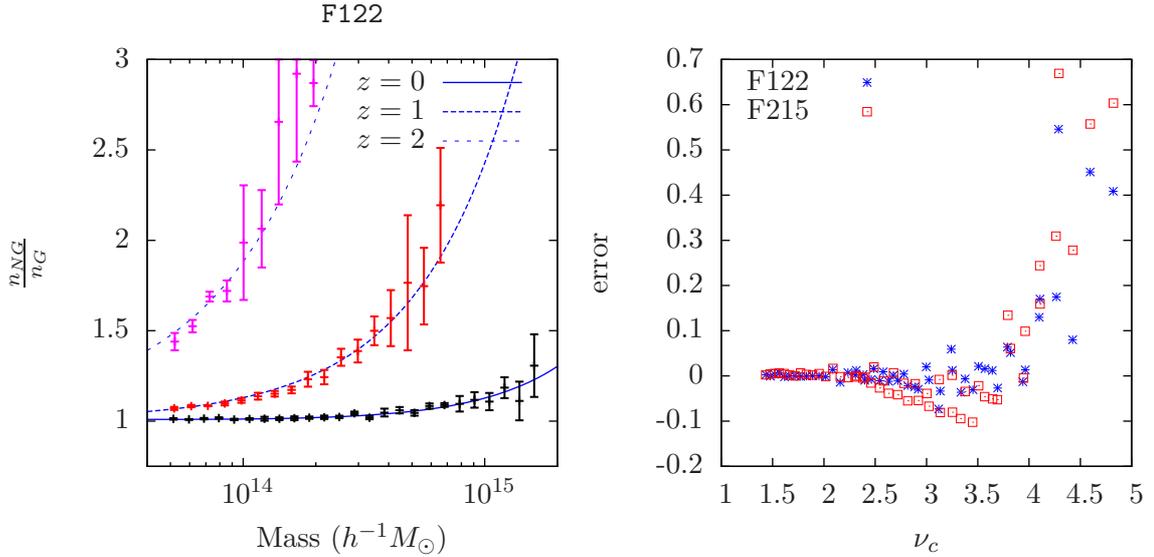

 \centering
 \begin{subfigure}[b]{0.48\textwidth}
  \centering
  \input{figures/eee.tex}
 \end{subfigure}
 \begin{subfigure}[b]{0.48\textwidth}
  \centering
  \input{figures/error.tex}
 \end{subfigure}
 \caption{Left: the simulation results and the mass function prediction Eq.(\ref{eq:feedmassfn}) for \texttt{F122} model: $\mathcal{M}_3=0.036$ ($\fNLeff=122$) with $f_2=1.021$. Right: the fractional error in the semi-analytic predictions for \texttt{F215} and \texttt{F122} models compared to simulation results. The plotted points include results from all three redshift values that we have looked at i.e. $z=0,1,2$.}
 \label{errorf}
\end{figure}

\begin{figure}
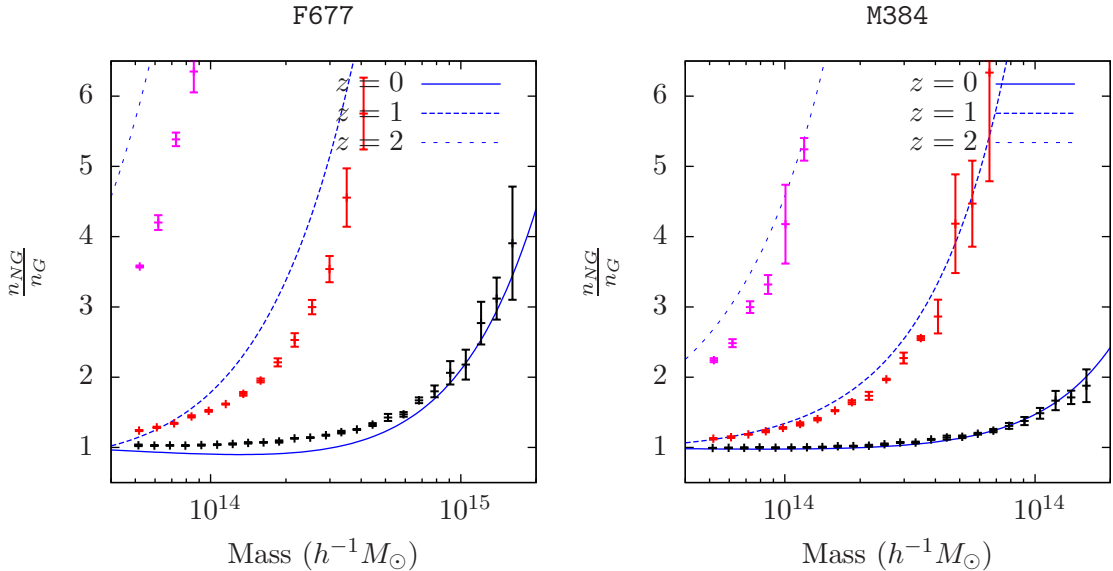

 \centering
 \begin{subfigure}[b]{0.48\textwidth}
  \centering
  \input{figures/cc.tex}
 \end{subfigure}
 \begin{subfigure}[b]{0.48\textwidth}
  \centering
  \input{figures/aa.tex}
 \end{subfigure}
 \caption{The simulation results and semi-analytic mass function prediction for: (i) left: a feeder simulation with  $\mathcal{M}_3\approx0.198$ ($f_2=1.16$), and (ii) right: a mixed scaling simulation with $\mathcal{M}_3\approx0.112$ ($f_2=1.06$).}
 \label{feedermixed}
\end{figure}

\begin{figure}
 \centering
 \begin{subfigure}[b]{\textwidth}
  \centering
  \input{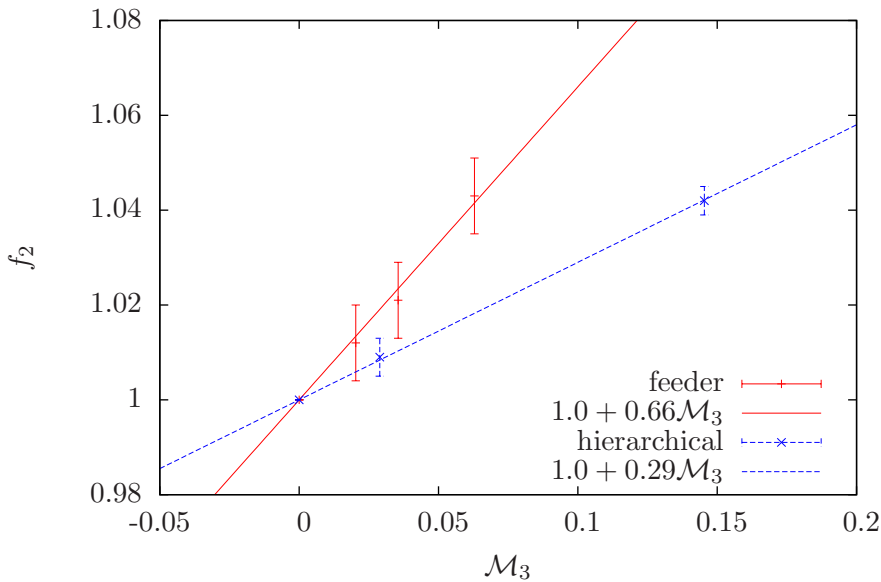}
 \end{subfigure}
 \caption{$\mathcal{M}_3$ dependence of $f_2$ defined in Eq.(\ref{eq:feedmassfn}) and Eq.(\ref{eq:hiermassfn}) for feeder and hierarchical mass functions respectively. The data points are best fit $f_2$ obtained from simulations and to obtain the best fit lines we require $f_2=0$ for $\mathcal{M}_3=0$.}
 \label{f2}
\end{figure}

The results for two feeder simulations are presented in the bottom panels of Figure \ref{mfnresults}. As discussed in Appendix \ref{app:erroranalysis}, the errors on our truncated feeder mass function are large compared to the hierarchical case with comparable $\mathcal{M}_3$. Also, the error becomes large at relatively small $\nu$ even for the feeder scaling case with $\mathcal{M}_3\approx0.03$. Therefore, based on our error evaluation, we do not expect that our feeder mass function describes the simulation results well for the more massive halos or at higher redshifts for which $\nu_c \gtrsim 3$. We find that the feeder mass function formula Eq.(\ref{eq:feedmassfn}) fits well our simulation results for \texttt{F70} and \texttt{F122} (see the left panel of Figure \ref{errorf} for the \texttt{F122} simulation results). Consistent with the error analysis of feeder mass function, Eq.(\ref{eq:feedmassfn}) fits to \texttt{F215} with $\mathcal{M}_3\approx0.063$ are not equally good. With a larger $\mathcal{M}_3 \approx 0.198$, the \texttt{F677} simulation is clearly not well fit by our truncated feeder mass function (see Figure \ref{feedermixed}). Finally, Figure \ref{mfnresults} also shows the difference between the simulations and Eq.(\ref{eq:feedmassfn}) {\it assuming $\delta_c=1.686$, $f_2=1$, and that the moments scale exactly as in} Eq.(\ref{eq:feederMn}). That is, the bottom-most panels illustrate how calibrating on simulations shifts the purely analytic expectations for the non-Gaussian mass function. The dominant effect among these three factors is that of $\delta_c$; change in $\delta_c$ affects the non-Gaussian mass functions starting at $\mathcal{O}(\mathcal{M}_3)$. On the other hand,  using different scalings of higher moments only change the expressions starting at $\mathcal{O}(\mathcal{M}_4)$ and the $f_2$ factors modify the mass functions at a few percent level at most (see Figure \ref{f2}).

In Figure \ref{errorf}, the right hand panel shows the fractional difference between the semi-analytic expression Eq.(\ref{eq:feedmassfn}) (with the cumulants $\mathcal{M}_n$ measured in the realizations and $\delta_c$, $f_2$ fit) and the simulation results for the two feeder models: \texttt{F122} and \texttt{F215}. We have plotted the fractional difference as a function of $\nu_c=\delta_c/\sigma_M$ and the plotted points include simulation results from all three redshifts $z=0,1,2$. The result can be interpreted as the error in the semi-analytic approach and qualitatively correlates with our analytic error analysis of the PDF: (i) the error at low $\nu_c$ is small, (ii) the error increases for higher $\nu_c$ but the error is smaller for smaller $\mathcal{M}_3$. 

On a different note, by looking at the non-Gaussian mass function results from simulations only, we can verify that the \texttt{F70} model is more non-Gaussian than the \texttt{H99} model (compare top and bottom plots in Figure \ref{mfnresults}), even though the skewness of \texttt{F70} is smaller than the skewness of \texttt{H99}. Similarly, we also see that the \texttt{F215} model has comparable amount of non-Gaussianity as the \texttt{H500} model. This verifies that the non-Gaussian mass function is sensitive to the total non-Gaussianity and the scaling of higher moments in the initial conditions.

The right hand panel of Figure \ref{feedermixed} is a mixed scaling simulation with $\mathcal{M}_3=0.112$, and approximately a third of the contribution to the $\mathcal{M}_3$ coming from the feeder component. For the theory curve we used both Eq.(\ref{eq:feedmassfn}) and Eq.(\ref{eq:hiermassfn}) for the corresponding $\mathcal{M}_3$ components but forced the same $f_2$. As with the case with other simulations containing feeder scaling, the high $\nu_c$ mass function results are not described well by the Edgeworth mass function.

All in all, our analysis of the truncation error and the Edgeworth fits to the mass functions from simulations were qualitatively consistent with each other. Hence, we find that the error evaluation of the PDF is a good indicator of the accuracy of the Edgeworth (or Petrov) mass function. However, to fit the simulations well we needed to rescale the analytic ratio of non-Gaussian to Gaussian mass function by an extra $\mathcal{M}_3$ dependent parameter, $f_2$, for both scalings. This rescaling seems reasonable to enforce the same total matter density regardless of level of non-Gaussianity. In Figure \ref{f2}, we plot our best fit $f_2$ values as a function of $\mathcal{M}_3$ for both scalings. We find a simple linear relation between $f_2$ and $\mathcal{M}_3$:
\begin{eqnarray}
 f_2^{\rm hier} &=& 1.0 + 0.29 \mathcal{M}_3 \\
 f_2^{\rm feed} &=& 1.0 + 0.66 \mathcal{M}_3 
\end{eqnarray}

\begin{figure}
 \centering
 \begin{subfigure}{0.48\textwidth}
  \centering
  \setlength{\unitlength}{0.0468bp}%
  \begin{picture}(4500,4500)
        \put(40,2500){\rotatebox{-270}{\makebox(0,0){\strut{}$\mathcal{M}_3$}}}%
        \put(2750,3750){hierarchical}
        \put(1700,4300){$\log_{10}\left(\frac{n_{NG}}{n_G}\right)$}
        \put(2400,60){$\nu_c$}
    \put(0,50){\includegraphics[scale=0.56]{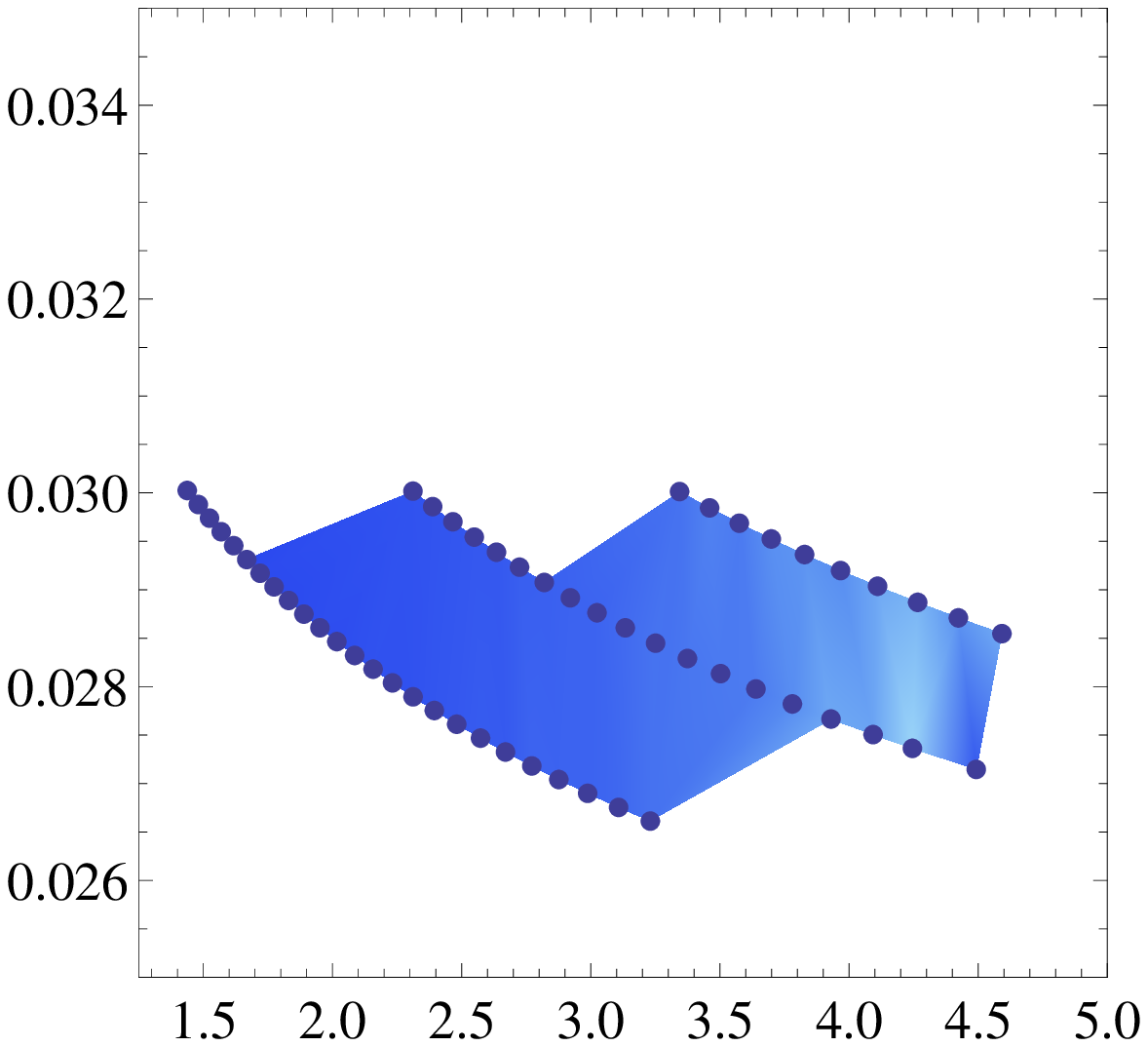}}%
  \end{picture}
 \end{subfigure}
 \begin{subfigure}{0.48\textwidth}
  \centering
  \setlength{\unitlength}{0.0468bp}%
  \begin{picture}(4500,4500)
    \put(3200,3750){feeder}
     \put(1700,4300){$\log_{10}\left(\frac{n_{NG}}{n_G}\right)$}
     \put(2400,60){$\nu_c$}
    \put(-150,35){\includegraphics[scale=0.76]{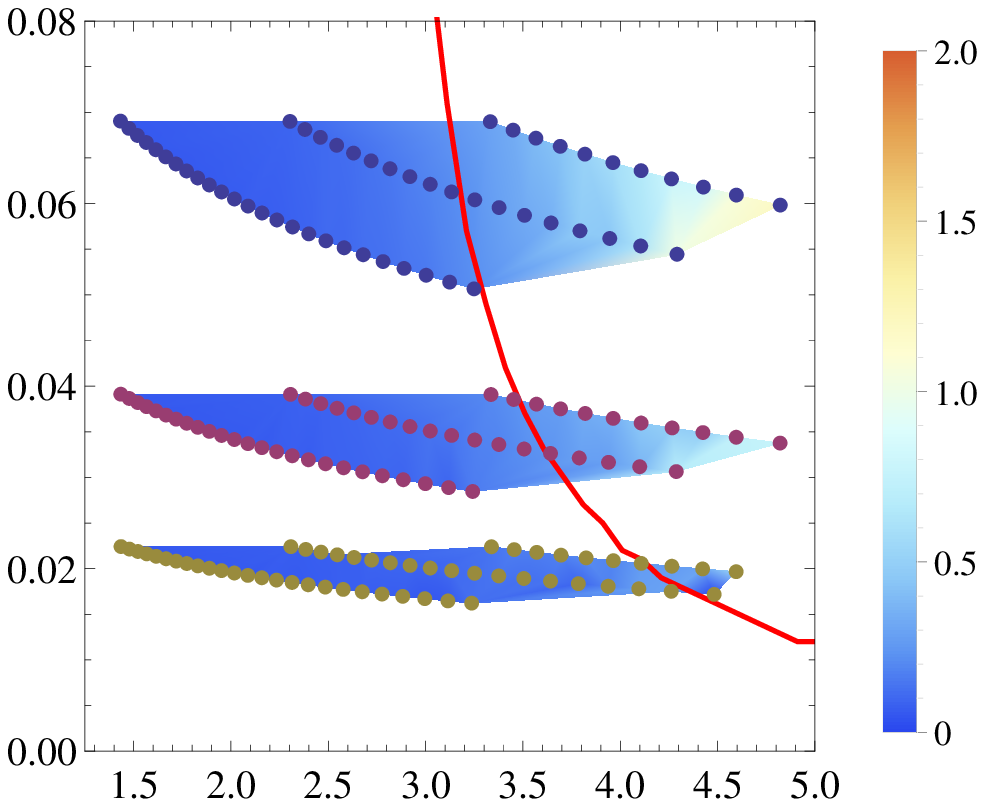}}%
  \end{picture}
 \end{subfigure}
 \caption{We plot results from our simulations for both hierarchical and feeder scalings as density plots. The quantity plotted is $\log_{10}(n_{NG}/n_G)_{\rm sim}$. The points overlayed on the plots are the data points used to obtain the density plots. The red solid line shows the maximum trusted $\mathcal{M}_3$ when allowing for 20\% error in the PDF from $\nu_c$ to some $\nu_{c, max}$. The maximum trusted $\mathcal{M}_3$ line for the hierarchical case lies outside of the plot range. See Appendix \ref{app:erroranalysis} for details of how these curves were obtained.}
 \label{nuhe6}
\end{figure}
\begin{figure}
 \centering
 \begin{subfigure}{0.48\textwidth}
  \centering
  \setlength{\unitlength}{0.0468bp}%
  \begin{picture}(4500,4500)
        \put(40,2500){\rotatebox{-270}{\makebox(0,0){\strut{}$\mathcal{M}_3$}}}%
        \put(2750,3750){hierarchical}
        \put(2000,4300){$R_{SA}$}
        \put(2400,60){$\nu_c$}
    \put(0,50){\includegraphics[scale=0.56]{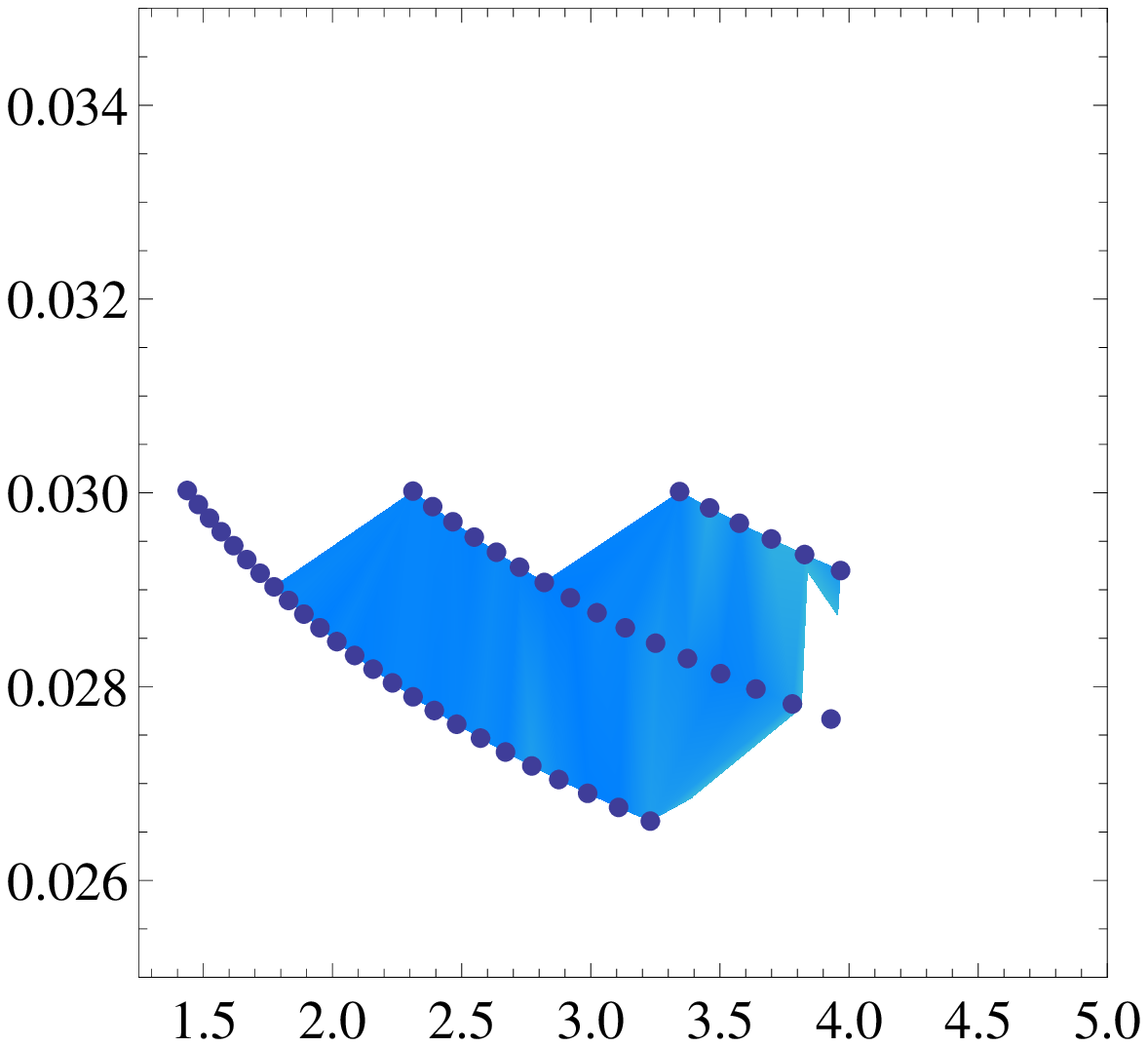}}%
  \end{picture}
 \end{subfigure}
 \begin{subfigure}{0.48\textwidth}
  \centering
  \setlength{\unitlength}{0.0468bp}%
  \begin{picture}(4500,4500)
    \put(3200,3750){feeder}
     \put(2000,4300){$R_{SA}$}
     \put(2400,60){$\nu_c$}
    \put(-150,35){\includegraphics[scale=0.76]{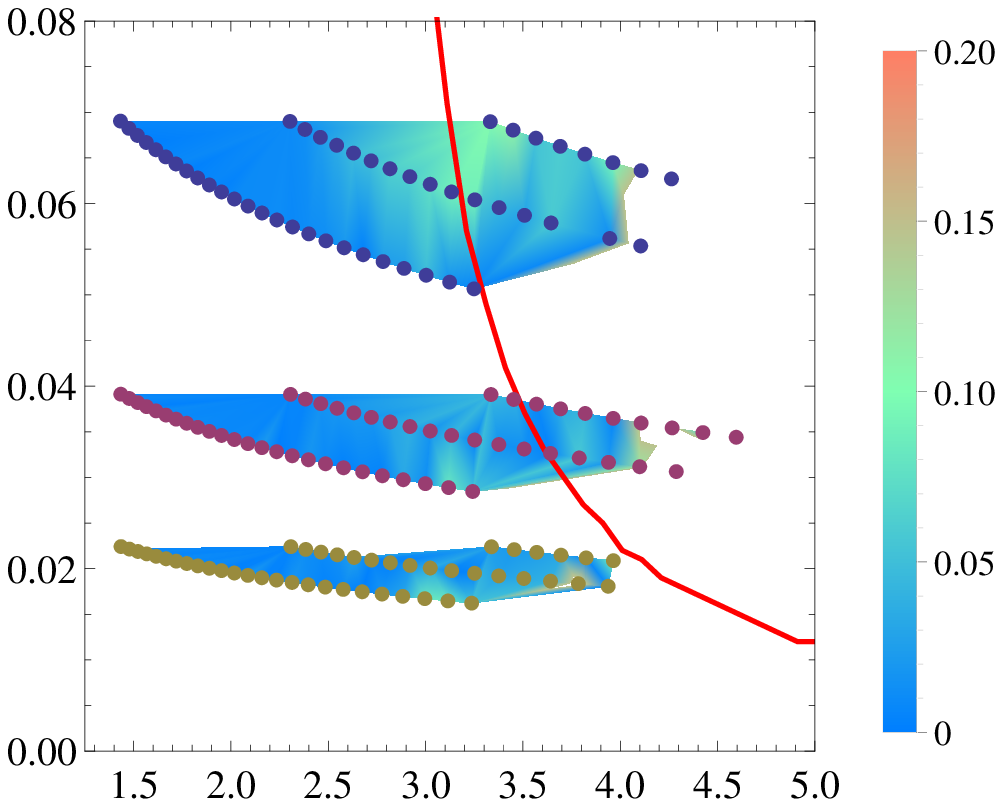}}%
  \end{picture}
 \end{subfigure}
 \caption{We plot the relative difference between the semi-analytic predictions (Eq.(\ref{eq:hiermassfn} or Eq.(\ref{eq:feedmassfn})) and the simulation results as a density plot. See Eq.(\ref{eq:RSA}) for the precise definition of the quantity plotted. We have omitted simulation results for which the uncertainty is larger than $20\%$. The important observation to note here is that the relative difference is quite small i.e. $\ll0.2$, below the solid red line.}
 \label{nuhe6error}
\end{figure}

Analytic suggestions for non-Gaussian mass functions have also been obtained using excursion set methods \cite{Maggiore:2009rx, Maggiore:2009hp, D'Amico:2010ta, DeSimone:2010mu, DeSimone:2011dn, Musso:2013pja, Achitouv:2013oea}. It would be interesting to compare the predictions for additional corrections to Eq.(\ref{nng}) from those methods to these simulations, particularly the results for $f_2$ shown in Figure \ref{f2}. 

To finish up our discussion of mass function results, in Figure \ref{nuhe6} we show the non Gaussian mass function results (from simulations) in a two dimensional density plot on a $\nu_c-\mathcal{M}_3$ plane, for both hierarchical and feeder scalings. We plot the quantity $\log_{10}\left(\left[n_{NG}/n_{G}\right]_{\rm sim}\right)$. In the same plots, we overlay the maximum $\mathcal{M}_3$ that we can trust the PDF \ref{eq:pdf}, $\mathcal{M}_{3,\rm max}$, as a function of $\nu_c$; see error analysis in Appendix \ref{app:erroranalysis} for details. In Figure \ref{nuhe6error}, we have plots similar to those in Figure \ref{nuhe6} but now we are plotting the relative difference of our mass function predictions (from \ref{eq:hiermassfn} and \ref{eq:feedmassfn}) from the simulation results. The quantity plotted is:
\begin{eqnarray}
 R_{SA}&=&\left|\frac{(n_{NG}/n_G)_{\rm sim} -(n_{NG}/n_{G})_{\rm semi-analytic}}{(n_{NG}/n_G)_{\rm sim}}\right|
 \label{eq:RSA}
\end{eqnarray}

These plots summarize our results for the mass function (for simulations with $\mathcal{M}_3<0.08$). Each isolated region in the density plots represents one simulation, which can be easily mapped to the exact simulation (in Table \ref{table:parameterspace}) by looking at the $\mathcal{M}_3$ value. Figure \ref{nuhe6} simply shows our simulation mass function results. But from Figure \ref{nuhe6error}, we we can see that the magnitude of relative difference of our semi-analytic mass functions to the simulation results is generally less than ten percent ($R_{SA}\lesssim0.1$) for the values of $\nu_c$ and $\mathcal{M}_3$ that are calculated to be trustworthy by evaluating the series truncation error of the PDF \ref{eq:pdf}. This is telling us that our semi-analytical mass function fits are consistent with the truncation error analysis.

\begin{figure}
 \centering
 \begin{subfigure}[b]{0.48\textwidth}
  \centering
  \input{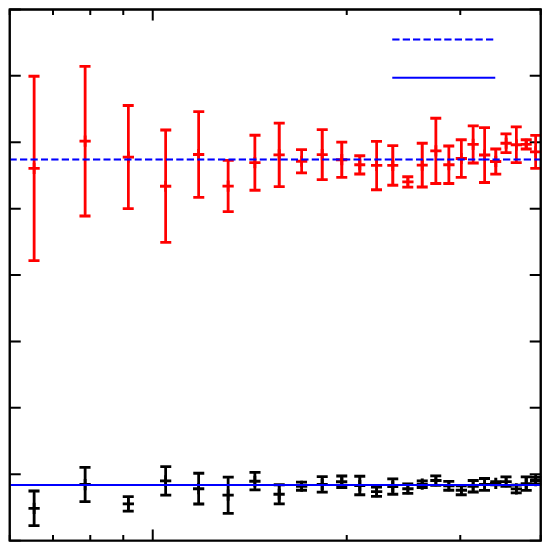}
 \end{subfigure}
 \begin{subfigure}[b]{0.48\textwidth}
  \centering
  \input{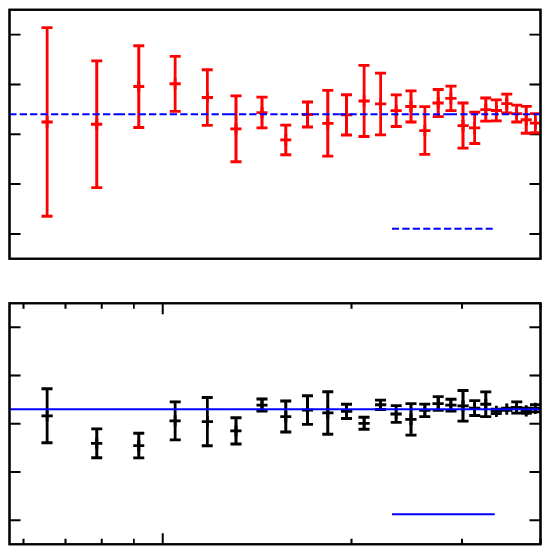}
 \end{subfigure}
 \caption{The bias (left) and stochasticity (right) at large scales for the Gaussian simulations at $z=0$ and $z=1$ using halos in the mass range $4.83 \times 10^{13} h^{-1} M_{\odot} \leq M \leq 9.55 \times 10^{13} h^{-1} M_{\odot}$. The values of large scale Gaussian bias and stochastic bias obtained through the fits can be found in Table \ref{table:gausbias}.}
\label{gbias}
\end{figure}

\paragraph{Bias and Stochasticity:} Next we present simulation results and fit to the analytical predictions for large scale bias and large scale stochasticity. Before considering the non-Gaussian results, we present the results from our Gaussian simulations in Figure \ref{gbias}. Both the bias and the stochastic bias (after the $1/\bar{n}$ shot noise correction to $P_{hh}(k)$) are constant for $k<0.04 \hMpc$, consistent with lowest order predictions for Gaussian fluctuations. Further, the large scale stochasticity parameter $\stochlss$ is predicted to approach unity for Gaussian initial conditions. Figure \ref{gbias} (right plot) is consistent with large scale stochasticity being constant for Gaussian initial conditions; however, the constant value is slightly greater than unity at $r^2_{\rm gaus}=1.03\pm0.02$ when halos in the mass range $4.83 \times 10^{13} h^{-1} M_{\odot} \leq M \leq 9.55 \times 10^{13} h^{-1} M_{\odot}$ are used at $z=0$. For our other $z=0$ samples, the value of $r^2_{\rm gaus}$ increased when using halos of larger mass. The values of the fitted parameters (bias and stochasticity) for our Gaussian simulations are summarized in Table \ref{table:gausbias}. We also note that the shot noise corrections are large for some of our halo samples. For example, the ratio of shot noise correction to the uncorrected halo power spectrum is largest for the halo sample at $z=2$ (Gaussian simulations) in Table \ref{table:gausbias}; at a reference wavenumber $k\approx0.02 \hMpc$, this ratio is $\approx 0.8$. Typically, this ratio is smaller ($\approx 0.3-0.5$) for the samples at smaller redshift in Table \ref{table:gausbias}. Also, note that for many of our non-Gaussian models, the fractional shot noise contribution decreases to $\approx 0.3$ even for the $z=2$ sample as there are more halos in the non-Gaussian samples at higher redshifts. Some progress has been made recently towards a better understanding and modeling of large scale stochasticity \cite{Baldauf2013}, beyond the usual shot noise correction, for Gaussian initial conditions. In this paper, however, we are focusing on the non-Gaussian effect only. 

\begin{table}
 \centering
 \begin{tabular}{|c|c|c|c|}
 \hline
  $z$ & Mass range of halos used & $b_g$ & $r^2_{\rm gaus}$ \\ \hline
  0 & $(1.93 \leq M \leq 3.85) \times 10^{14} h^{-1} M_{\odot}$& $3.36\pm0.09$& $1.13 \pm 0.06$ \\ 
   & $(0.965 \leq M \leq 1.92) \times 10^{14} h^{-1} M_{\odot}$ & $2.57\pm0.04$ & $1.08 \pm 0.03$ \\ 
   & $(4.83 \leq M \leq 9.55) \times 10^{13} h^{-1} M_{\odot}$ & $1.92\pm0.04$ & $1.03\pm 0.02$ \\ \hline
  $1$ & $(0.965 \leq M \leq 1.92) \times 10^{14} h^{-1} M_{\odot}$ & $5.83\pm0.24$ & $1.01\pm0.10$ \\
  & $(4.83 \leq M \leq 9.55) \times 10^{13} h^{-1} M_{\odot}$& $4.37 \pm 0.11$ & $1.04 \pm 0.05$ \\ \hline
  $2$ & $(4.83 \leq M \leq 9.55) \times 10^{13} h^{-1} M_{\odot}$ & $8.68\pm0.44$ & $0.84 \pm 0.29$ \\ \hline
 \end{tabular}
\caption{Large scale bias and stochasticity for simulations with Gaussian initial conditions. Simple chisquare fitting was performed to obtain the best fit values. The errors in $b_g$ and $\stochlss$ are computed such that the reduced chisquare increases by unity when adding the error to the best fit values. We will use the same procedure to obtain best fit values and the corresponding error for our non-Gaussian bias and stochasticity results.}
\label{table:gausbias}
\end{table}

\begin{figure}
\begin{subfigure}[b]{0.48\textwidth}
  \centering
  \input{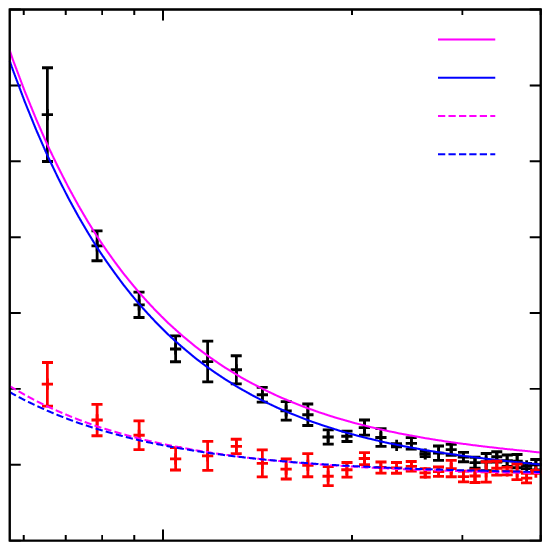}
 \end{subfigure}
 \begin{subfigure}[b]{0.48\textwidth}
  \centering
  \input{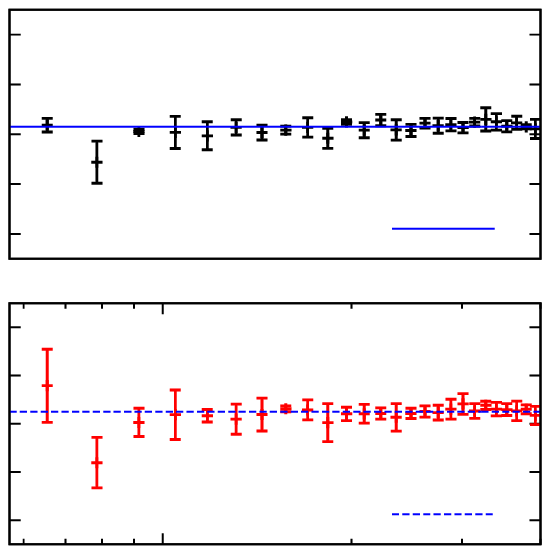}
 \end{subfigure}
 \caption{Left: the simulation result for bias $P_{hm}/P_{mm}$ at large scales for the hierarchical simulations at $z=0$ using halos in the mass range $9.65 \times 10^{13} h^{-1} M_{\odot} \leq M \leq 1.92 \times 10^{14} h^{-1} M_{\odot}$ and the corresponding best fit using Eq.(\ref{eq:hmbiasfinal}). For each set of simulation data, we also include the curve obtained by using $b_g$ from Gaussian simulation instead of the fitted $b_\psi$. For the case of \texttt{H500} shown in the figure, $b_\psi=1.86\pm0.03$ compared to $b_g=1.92\pm0.04$. Right: the corresponding large scale stochasticity simulation results Eq.(\ref{eq:stochasticity}) and the best fit constant values.}
 \label{hbias}
\end{figure}

We have derived expressions for $P_{hm}(k)$ and $P_{hh}(k)$ for our two source model in Appendix \ref{app:bias}. These expressions written in terms of the total matter power spectrum $P_{mm}(k)=\alpha(k)^2 P_{\Phi}(k)$ and the constant halo bias coefficients ($b_\phi$ and $b_\psi$) for the two independent fields $\phi$ and $\psi$ are:
\begin{eqnarray}
 \frac{P_{hm}(k)}{P_{mm}(k)}&=&\frac{b_\phi(1-q)}{1+\ftNL^2 I_1(k)q\mathcal{P}_{\psi,G}}+\left(b_\psi+2\delta_c(b_\psi-1)\frac{\ftNL}{\alpha(k)}\frac{\sigma_{R,\psi}^2}{\sigma_R^2}\right) \left(\frac{q+\ftNL^2I_1(k)q\mathcal{P}_{\psi,G}(k)}{1+\ftNL^2I_1(k)q\mathcal{P}_{\psi,G}}\right)\; \nonumber \\
 \label{eq:hmbiasfinal}
\end{eqnarray}
and,
\begin{eqnarray}
 \frac{P_{hh}(k)}{P_{mm}(k)} &=& \frac{b_\phi^2 (1-q)}{1+\ftNL^2 I_1(k)q\mathcal{P}_{\psi,G}} + \left(b_\psi+2\delta_c(b_\psi-1)\frac{\ftNL}{\alpha(k)}\frac{\sigma_{R,\psi}^2}{\sigma_R^2}\right)^2\left( \frac{q+\ftNL^2I_1(k)q\mathcal{P}_{\psi,G}(k)}{1+\ftNL^2 I_1(k)q\mathcal{P}_{\psi,G}}\right)\; \nonumber \\
 \label{eq:hhbiasfinal}
\end{eqnarray}

For each model (listed in Table \ref{table:parameterspace}) we have measured the auto and cross correlations in the matter and halo fields, using halos in the same mass bins and redshifts shown in Table \ref{table:gausbias}. To check how well the analytic expressions above fit the simulation results, we measured the bias coefficients $b_\phi$ and $b_\psi$ for each case by cross-correlating the halo density field for each sample with (i) the $\phix$ part of the linear density field, and (ii) the $\psix+\ftNL\psix^2$ part of the linear density field. For the correlation with the Gaussian field, we expect a constant large scale bias $b_\phi$. For the second case we expect a scale-independent piece and a scale dependent term. The expression for this bias is obtained using Eq.(\ref{eq:deltah}) by cross correlating $\delta_h$ and $\delta_{\psi, \rm NG}$, to get:
\begin{eqnarray}
 \frac{P_{h\psi}}{P_{\psi\psi}} &=& b_\psi + \frac{2\delta_c (b_\psi-1)\ftNL}{\alpha(k)} \frac{\sigma_{R,\psi}^2}{\sigma_R^2}
\end{eqnarray} 

Equations (\ref{eq:hmbiasfinal}) and (\ref{eq:hhbiasfinal}) clearly give the expected result for the Gaussian case in the $q=0$ limit; in this case only the $\phi$ field contributes to the initial density field. We obtain the single field non-Gaussian (hierarchical, feeder, or mixed) limit for $q=1$, in which case there is no contribution from the $\phi$ field to the initial density field. Further, in case of single field hierarchical models,

\begin{eqnarray}
 \left(\frac{1 + \ftNL^2 \mathcal{P}_{\psi,G}(k) I_1(k)}{1+\ftNL^2 q \mathcal{P}_{\psi,G}(k) I_1(k)} \right) &=& 1 \;, {\rm  and}\; \frac{\sigma_{R,\psi}^2}{\sigma_R^2}\approx q
\end{eqnarray}
and we recover the known results for the local ansatz. Using the general expressions in Eq.(\ref{eq:hmbiasfinal}) and Eq.(\ref{eq:hhbiasfinal}) and the best fit bias parameters for each source field, we will compare the semi-analytical predictions for the bias and the stochasticity parameter $\stochlss$ to our simulation results. 

\begin{figure}
 \centering
 \begin{subfigure}[b]{0.48\textwidth}
  \centering
  \input{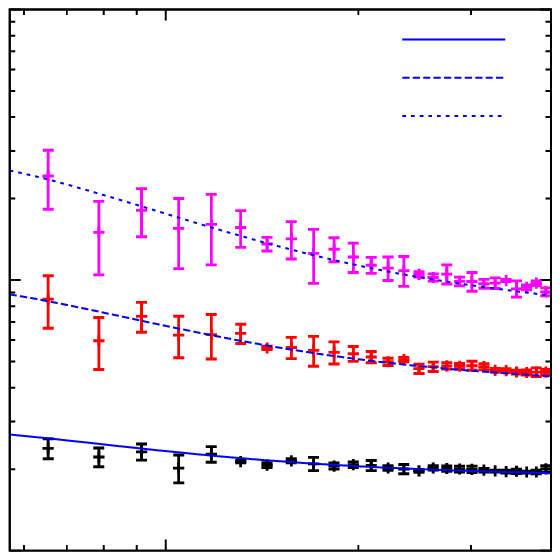}
 \end{subfigure}
 \begin{subfigure}[b]{0.48\textwidth}
  \centering
  \input{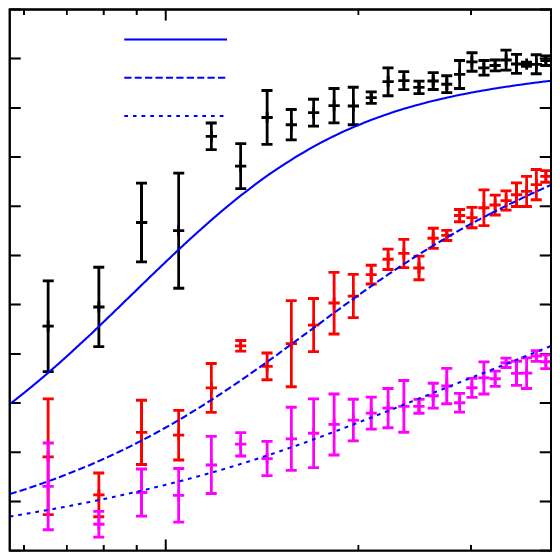}
 \end{subfigure}
 \begin{subfigure}[b]{0.48\textwidth}
  \centering
  \input{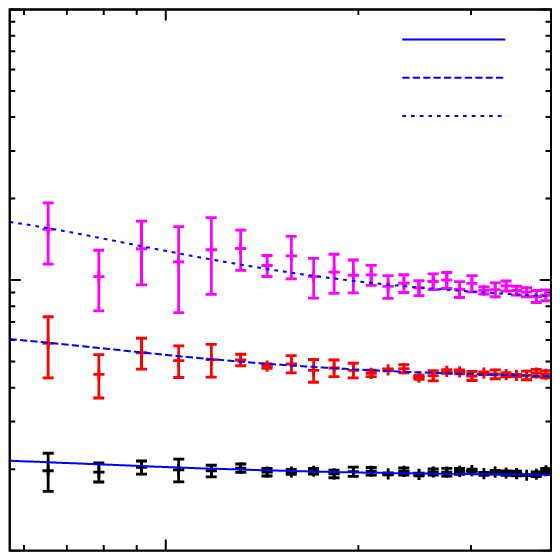}
 \end{subfigure}
 \begin{subfigure}[b]{0.48\textwidth}
  \centering
  \input{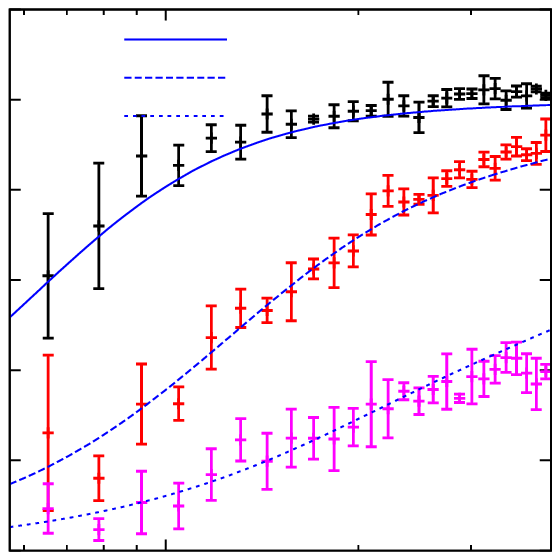}
 \end{subfigure}
 \caption{\textit{Top}: the bias $P_{hm}/P_{mm}$ (left) and stochasticity $\stochlss$ (right) at large scales for the feeder simulation \texttt{F677} at $z=0, 1, 2$ using halos in the mass range $4.83 \times 10^{13} h^{-1} M_{\odot} \leq M \leq 9.55 \times 10^{13} h^{-1} M_{\odot}$. The analytical curves for the bias is obtained using Eq.(\ref{eq:hmbiasfinal}) for which the bias coefficients $b_\psi$ and $b_\phi$ are measured by cross-correlating each halo sample with the linear density field contribution from the $\phi$ and $\psi$ fields respectively. The analytical curves for the stochasticity is obtained using Eq.(\ref{eq:hmbiasfinal}) and Eq.(\ref{eq:hhbiasfinal}) in Eq.(\ref{eq:stochasticity}) using the same $b_\phi$ and the same $b_\psi$ as the corresponding bias curve on the left plot. \textit{Bottom}: same as top but for the model \texttt{F215} using the same halo samples. All the analytical curves for the bias are consistent with the simulation results. For the stochasticity  $\stochlss$, the most discrepant case in the above plots (the \texttt{F677}, $z=0$ sample) is off by about five percent.}
\label{fbias}
 \end{figure}

Now, let us discuss the single field hierarchical scenario first. Note that we use $\delta_c=1.46$, the best fit $\delta_c$ from our mass function fits. We find that a different best fit $b_\psi$ is preferred compared to the corresponding Gaussian bias $b_g$ measured from Gaussian simulations. This can be clearly seen in Figure \ref{hbias}, especially in the case of \texttt{H500}. We checked this to be true for other halo samples listed in Table \ref{table:gausbias}. In general, we find the best fit $b_\psi$ is less than the corresponding $b_g$; this is consistent with the picture that bias decreases as mass function increases \cite{Desjacques2009}. From Figure \ref{hbias} (right plot), we can also verify that the single field non-Gaussian cases do not produce excess stochasticity than the Gaussian case as predicted. For the two more massive halos samples at $z=0$, for which the Gaussian stochasticity itself deviates from unity, we find similar level of deviation from unity in the non-Gaussian case. 
 
\begin{figure}
 \centering
 \begin{subfigure}[b]{0.48\textwidth}
  \centering
  \input{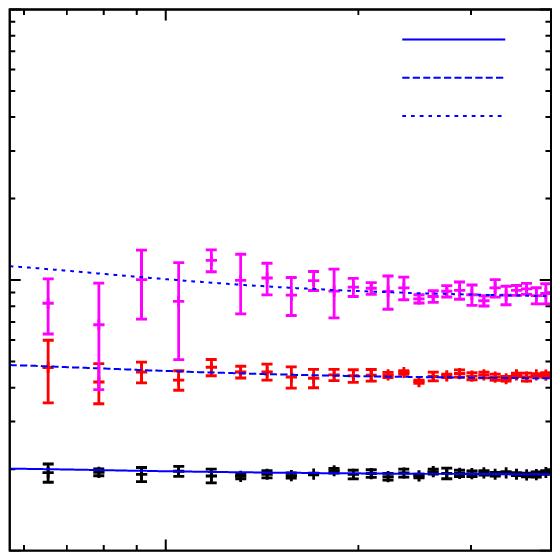}
 \end{subfigure}
 \begin{subfigure}[b]{0.48\textwidth}
  \centering
  \input{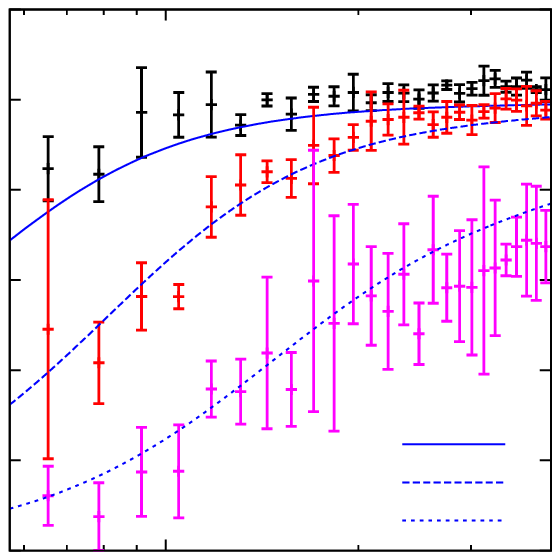}
 \end{subfigure}
 \begin{subfigure}[b]{0.48\textwidth}
  \centering
  \input{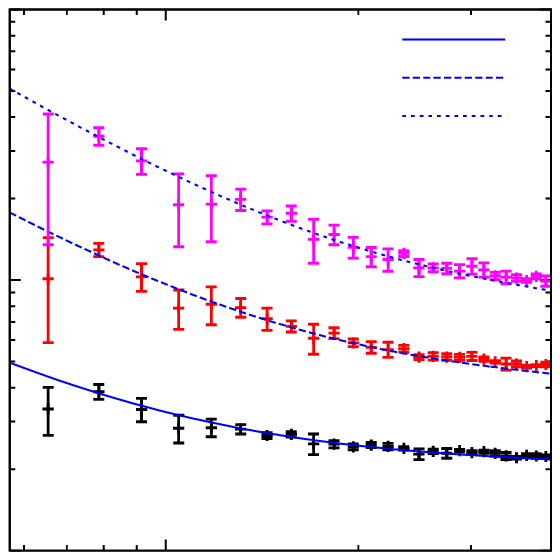}
 \end{subfigure}
 \begin{subfigure}[b]{0.48\textwidth}
  \centering
  \input{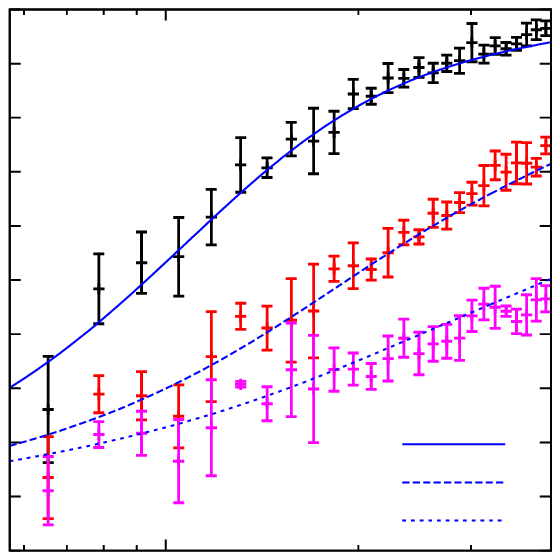}
 \end{subfigure}
 \caption{\textit{Top}: the bias $P_{hm}/P_{mm}$ (left) and stochasticity $\stochlss$ (right) at large scales for the feeder simulation \texttt{F70} at $z=0, 1, 2$ using halos in the mass range $4.83 \times 10^{13} h^{-1} M_{\odot} \leq M \leq 9.55 \times 10^{13} h^{-1} M_{\odot}$. The analytical curves for the bias are obtained using Eq.(\ref{eq:hmbiasfinal}) for which the bias coefficients $b_\psi$ and $b_\phi$ are measured by cross-correlating each halo sample with the linear density field contribution from the $\phi$ and $\psi$ fields respectively. The analytical curves for the stochasticity are obtained using Eq.(\ref{eq:hmbiasfinal}) and Eq.(\ref{eq:hhbiasfinal}) in Eq.(\ref{eq:stochasticity}) using the same $b_\phi$ and the same $b_\psi$ as the corresponding bias curve on the left plot. \textit{Bottom}: same as top but for the model \texttt{M997} using the same halo samples.}
\label{fmbias}
 \end{figure}

In Figure \ref{fbias}, we show bias and stochasticity results for two of our feeder cases. We find that for the cases in which the corresponding $r^2_{\rm gaus}$ is consistent with unity, the simulation results for the stochasticity are described quite well by the analytic expression. In Figure \ref{fmbias}, we show another example of a feeder case \texttt{F70} and an example for a mixed case \texttt{M997}. In Table \ref{table:phipsibias}, we list the values $b_\psi$ and $b_\phi$ used in the plots.

\begin{table}
 \centering
 \begin{tabular}{|c|cc||cc||cc|}
 \cline{2-7}
  \multicolumn{1}{c}{ } & \multicolumn{2} {|c||} {z=0} & \multicolumn{2} {|c||} {z=1} & \multicolumn{2} {|c|} {z=2} \\ \cline{2-7}
   \multicolumn{1}{c|}{}& $b_\phi$ & $b_\psi$ & $b_\phi$ & $b_\psi$ & $b_\phi$ & $b_\psi$ \\ \hline
  F677 &  $1.76\pm0.05$ & $1.85\pm0.05$ & $3.34\pm0.19$ & $4.15\pm0.09$ & $4.49\pm0.31$ &$9.24\pm0.17$ \\
  F215 &  $1.83\pm0.05$ & $2.08\pm0.13$ & $3.96\pm0.14$ & $5.04\pm0.10$ & $6.36\pm0.63$ &$14.36\pm0.48$ \\
  F70 &  $1.88\pm0.04$ & $2.25\pm0.23$ & $4.18\pm0.11$ & $5.09\pm0.23$ & $7.88\pm0.55$ &$15.08\pm0.96$ \\
  M997 &  $1.87\pm0.08$ & $1.95\pm0.06$ & $2.95\pm0.09$ & $3.83\pm0.06$ & $4.16\pm0.46$ &$7.18\pm0.11$ \\
  \hline
 \end{tabular}
\caption{Values of the bias coefficients $b_\phi$ and $b_\psi$ measured by cross correlating the halo density field with the corresponding $\phi$ and $\psi$ components in the linear density field. In all redshifts $z=0,1,2$ listed above, the mass range for the halo samples used was $4.83 \times 10^{13} h^{-1} M_{\odot} \leq M \leq 9.55 \times 10^{13} h^{-1} M_{\odot}$. These same halo samples are used for the bias and stochasticity plots shown in Figures \ref{fbias} and \ref{fmbias}.}
\label{table:phipsibias}
\end{table}
 
We have only shown bias and stochasticity results for three samples out of the six samples listed in Table \ref{table:gausbias}. Let us comment on the results obtained for the samples for which results are not presented here. First, we find that the bias $P_{hm}/P_{mm}$ results agree with the analytic prediction for all the samples. Similarly, the stochasticity $\stochlss$ for the $z=1$ sample with mass range $9.65 \times 10^{13} h^{-1} M_{\odot} \leq M \leq 1.92 \times 10^{14} h^{-1} M_{\odot}$ has excellent fits to the simulation data for all the feeder and mixed models. However, we find that the stochasticity $\stochlss$ results for the two samples at $z=0$ with mass ranges: (i) $(0.965 \leq M \leq 1.92) \times 10^{14} h^{-1} M_{\odot}$ and (ii) $(1.93 \leq M \leq 3.85) \times 10^{14} h^{-1} M_{\odot}$ show deviations from the analytic predictions, roughly at the same level as the deviation shown by corresponding Gaussian $r^2_{\rm gaus}$ from unity (at $\approx 10\%$ level).

Reference \cite{Smith2011} found that the stochastic bias predictions for one case of a two field model were off by roughly a factor of $0.7$. For comparison, we define their $b_0$ in our notation:
\begin{eqnarray}
 b_0 &=&\frac{(1-q)b_\phi + q b_\psi}{1+\ftNL^2 q \mathcal{P}_{\psi,G}(k) I_1(k)}
\end{eqnarray}
For their particular model, $b_0 \approx \frac{1}{2}(b_\phi+b_\psi)$, and their model assumes $b_0=b_\phi=b_\psi$. However, our results indicate that we may have to relax this assumption in general two source non-Gaussian scenarios. We expect better match between simulation results and peak background split calculation in their work too, once $b_\psi$ and $b_\phi$ are measured separately, in addition to accounting for the $\mathcal{O}(\ftNL^3)$ term in the bispectrum, and the $\mathcal{O}(\ftNL^4)$ term in the trispectrum.

\section{Conclusion}
\label{sec:conclusion}
In this paper, we have reported results for the mass function of massive halos and scale dependent bias from N-body simulations using a generalized two field model of primordial non-Gaussianity. The two field ansatz is used to generate initial conditions where the higher moments $\mathcal{M}_{n}$, $n>3$ are more important than in the standard local ansatz, given the same value of $\mathcal{M}_3$ as the single field local ansatz case. That this can be done is not mathematically surprising, but we have considered a range of scalings that are very natural from particle physics models of inflation. We have shown that in using large scale structure data to constrain the primordial fluctuations, assumptions about the scaling of higher moments should be explicitly stated. In addition, we have shown that in scenarios with more than one source for the density fluctuations it is important to allow independent bias coefficients {\it for each source}.

Our simulations show that the Petrov expansion gives a good approximation to the non-Gaussian mass function when the amount of non-Gaussianity is small enough. The criteria for ``small enough" depends on the dimensionless skewness $\mathcal{M}_3$ {\it and} the scaling of moments. We have verified that for the same level of skewness, the feeder scaling is more non-Gaussian than the hierarchical scaling by a straightforward comparison of simulation outputs of the two cases. For the parameter space we probed, we were able to show that the truncation error evaluation of the Petrov PDF, Eq.(\ref{eq:pdf}), correlated well with the degree to which the truncated non-Gaussian mass function fitted the simulation results. This gives us further confidence in the use of the truncated Petrov (Edgeworth) mass function for various analyses such as that of \cite{Shandera2012, Shandera2013} to put constraints on primordial non-Gaussianity using number counts of objects. We make progress towards calibrating the mass function formulae Eq.(\ref{eq:hiermassfn}) and Eq.(\ref{eq:feedmassfn}) for both scalings by using an extra parameter $f_2$, in addition to verifying that a reduced $\delta_c$ is preferred, which is consistent with previous simulation studies of non-Gaussian mass functions with non-Gaussianity of local type and hierarchical scaling of higher moments. The effect this calibration might have on previous analysis of cluster constraints is illustrated in the bottom panels of Figure \ref{mfnresults}.

Similarly, the peak-background-split calculations are good fits to the non-Gaussian bias results ($P_{hm}(k)/P_{mm}(k)$) from simulations---both for the hierarchical models (previously done a number of times) and for the feeder models. Moreover, we also presented results for large scale stochastic bias from our simulations which, in addition to the halo-matter bias, also depends on $P_{hh}(k)/P_{mm}(k)$. We found that, for two source scenarios, both the bias and stochastic bias calculations work well once we allow for different bias coefficients for the two independent fields. These bias coefficients, namely $b_\phi$ and $b_\psi$ in our notation, were measured by cross-correlating the halo density field with the linear density field contributions from $\phi$ and $\psi$ fields separately. The analytical calculations deviated from the simulation results for the stochastic bias only for halo samples whose values of stochasticity were inconsistent with unity in the Gaussian simulations. Therefore, for these halo samples, we expect the need to account for other contributions to the stochasticity that have to be taken into account even in the Gaussian case. While we have not investigated this issue in detail, additional studies would be worthwhile since a better understanding of stochasticity is useful for cosmological applications of galaxy surveys.

\paragraph{Acknowledgment:} This work is supported by the National Aeronautics and Space Administration under Grant No. NNX12AC99G issued through the Astrophysics Theory Program. This work used the Extreme Science and Engineering Discovery Environment (XSEDE), which is supported by National Science Foundation grant number OCI-1053575. The simulations for the project were performed using computing resources at the Texas Advanced Computing Center (TACC), The University of Texas at Austin through XSEDE grant AST130053 and at the Penn State Research Computing and Cyberinfrastructure (RCC). S. A. would like to thank Xinghai Zhao for suggestions on setting up Gadget simulations.

\appendix
\section{Integrals for $\langle \delta_R^n\rangle_c$}
\label{integrals}
Here we list some of the integrals for $\langle \delta_R^n \rangle_c = \langle \delta_R^n \rangle_1 + \langle \delta_R^n\rangle_2$, where the subscripts $1$ and $2$ are for the $\mathcal{O}(\ftNL^{n-2}$) and $\mathcal{O}(\ftNL^n$) terms respectively.
\begin{eqnarray}
 \langle \delta_R^2\rangle_1 &=& \frac{1}{q} \int \frac{d^3\vec{k}}{\tpc} \alpha(k)^2 W_R(k)^2  P_{\psi}(k) = \frac{1}{q} \int \frac{dk}{k} \alpha(k)^2 W_R(k)^2 \mathcal{P}_{\psi}(k)  \\
 \langle \delta_R^2 \rangle_2 &=& 2 \ftNL^2 \int \frac{d^3\vec{k}}{\tpc} \int \frac{d^3\vec{p}}{\tpc} \alpha(k)^2 W_R(k)^2 P_{\psi}(p) P_{\psi}(|\vec{p}-\vec{k}|) \nonumber \\
 &=& \ftNL^2 \int k^2 dk \int dp \dmuint \left[\alpha(k)^2 W_R(k)^2 \frac{\mathcal{P}_{\psi}(p) \mathcal{P}_{\psi}(|\vec{p}-\vec{k}|)}{p |\vec{p}-\vec{k}|^3} \right]
\end{eqnarray}
The $\langle \delta_R^3\rangle$ terms are:
\begin{eqnarray}
 \langle \delta_R^3 \rangle_1 &=& 6 \ftNL \int \frac{d^3\vec{k_1}}{\tpc} \int
\frac{d^3 \vec{k_2}}{\tpc} (\alpha W)_{1,2,12} P_{\psi}(k_1) P_{\psi}(k_2) \nonumber\\
&=& 6  \ftNL \left[ \frac{1}{2} \int \frac{dk_1}{k_1} \int \frac{dk_2}{k_2}
\dmuint \left[ (\alpha W)_{1,2,12} \mathcal{P}_{\psi}(k_1) \mathcal{P}_{\psi}(k_2)\right] \right]
\end{eqnarray}
where,  $(\alpha W)_{1,2,12}=\alpha(k_1)W(k_1)\alpha(k_2)W(k_2)\alpha(|\vec{k_1}-\vec{k_2}|)W(|\vec{k_1}-\vec{k_2}|)$.
\begin{eqnarray}
 \langle \delta_R^3\rangle_2 &=& 8 \ftNL^3 \int \frac{d^3\vec{k_1}}{\tpc} \int
\frac{d^3 \vec{k_2}}{\tpc} (\alpha W)_{1,2,12} \int \frac{d^3\vec{p}}{\tpc}
P_{\psi}(p) P_{\psi}(|\vec{k_1}-\vec{p}|) P_{\psi}(|\vec{k_2}+\vec{p}|) \nonumber \\
&=&\frac{\ftNL^3}{2 \pi^2} \int k_1^2 dk_1 \int
k_2^2 dk_2 \dmuint_2 \int_0^{2\pi} d\phi_2 (\alpha W)_{1,2,12} \int \frac{dp}{p} \nonumber 
\\ & & \dmuint
\int_0^{2\pi} d\phi \left[ \frac{\mathcal{P}_{\psi}(p) \mathcal{P}_{\psi}(|\vec{k_1}-\vec{p}|) \mathcal{P}_{\psi}(|\vec{k_2}+\vec{p}|)}{|\vec{k_1}-\vec{p}|^3 |\vec{k_1}+\vec{p}|^3} \right]
\end{eqnarray}
where, $\phi$ is the angle between $\vec{k_1}$ and $\vec{p}$ and $\phi_2$ is the angle between $\vec{k_2}$ and $\vec{p}$.

\begin{figure}
 \centering
 \begin{subfigure}[b]{\textwidth}
  \centering
  \input{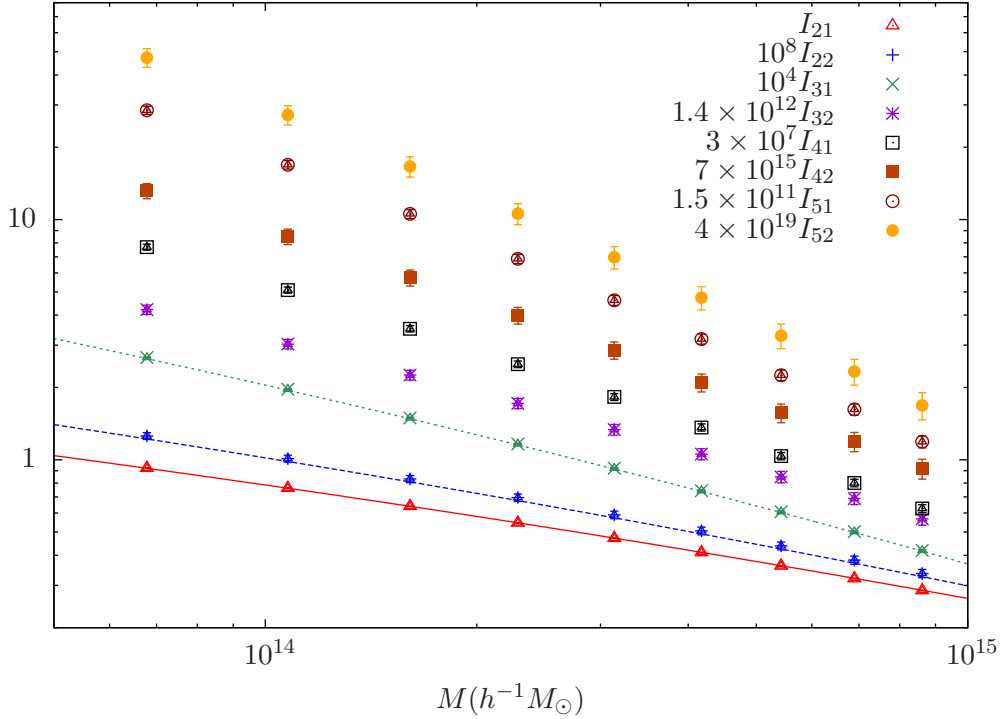}
 \end{subfigure}
 \caption{Data points are the cumulant estimates from the Monte-Carlo approach. See the text and specifically Eq.(\ref{eq:Iijdef}) and E.(\ref{eq:deltadef}) for the definition of $I_{ij}(M)$ plotted here. Since these integrals cover a huge range of values, we are multiplying by appropriate factors as shown in the plot for clarity. The dashed lines are the corresponding results from direct numerical integrations.}
 \label{smoothedmoments}
\end{figure}

In the above integrals, all the $k, p$ integrals go from $k_{min}$ to $k_{max}$ and the divergent pieces in the integrands are set to be greater than $k_{min}$; for example in $\langle \delta_R^2\rangle_2$ we set $|\vec{p}-\vec{k}|\geq k_{min}$---otherwise the integral diverges. The $\langle \delta^2_R \rangle_1$, $\langle \delta^2_R\rangle_2$ and $\langle \delta^3_R\rangle_1$ integrations were evaluated using Mathematica 9's \texttt{Adaptive MonteCarlo} numerical integration routine. To estimate the higher order cumulants, we used the Monte-Carlo approach described in Appendix A of \cite{Loverde2011}. When applying this method to the cumulants for which numerical integration was performed, the results agree. Following \cite{Loverde2011}, we simulate a Gaussian initial curvature $\Phi_G$ and define two fields by
\begin{eqnarray}
 \delta_{M, \Phi_G}(\vec{k})&=& W_M(k)\alpha(k)\int d^3\vec{x} e^{-i\vec{k}\cdot\vec{x}} \Phi_G(\vec{x}) \nonumber \\
 \delta_{M,\Phi_G}^{*}(\vec{k})&=& W_M(k)\alpha(k)\int d^3\vec{x}e^{-i\vec{k}\cdot\vec{x}} \Phi_G(\vec{x})^2
 \label{eq:deltadef}
\end{eqnarray}
Then the integral estimates are given by,
\begin{eqnarray}
 I_{21}(M) &=& \langle \delta_{M,\Phi_G}(\vec{x})^2\rangle \nonumber \\
 I_{22}(M) &=& \langle \delta_{M,\Phi_G}^*(\vec{x})^2\rangle \nonumber \\
 I_{31}(M) &=& 3 \langle \delta_{M, \Phi_G}(\vec{x})^2 \delta_{M, \Phi_G}^{*}(\vec{x})\rangle \nonumber \\
 I_{32}(M) &=& \langle \delta_{M, \Phi_G}^*(\vec{x})^3\rangle \nonumber \\
 I_{41}(M) &=& 6 \langle \delta_{M, \Phi_G}(\vec{x})^2 \delta_{M, \Phi_G}^*(\vec{x})^2\rangle \nonumber \\
 I_{42}(M) &=& \langle \delta_{M, \Phi_G}^*(\vec{x})^4\rangle \nonumber \\
 I_{51}(M) &=& 10 \langle \delta_{M, \Phi_G}(\vec{x})^2 \delta_{M, \Phi_G}^*(\vec{x})^3\rangle \nonumber \\
 I_{52}(M) &=& \langle \delta_{M,\Phi_G}^*(\vec{x})^5 \rangle
 \label{eq:Iijdef}
\end{eqnarray}

In Figure \ref{smoothedmoments}, we plot these values obtained through four Monte-Carlo realizations in a box of size $L=2400 \Mpch$ and same cosmological parameters as that of our N-body simulations. The error reported is $1\sigma$ variation from four realizations. To get the value of $\langle \delta_M^n\rangle_j$ ($j=1,2$) for each of our models parametrized by $q$ and $\ftNL$, $I_{nj}(M)$ should simply be multiplied by an appropriate factor which depends on our model parameters $q$ and $\ftNL$. For example, $\langle \delta_M^2\rangle_1 = \frac{1}{q} \frac{A_\psi}{A_{\Phi}} I_{21}(M)$ and $\langle \delta^3_M\rangle_1 = \left(\frac{A_{\psi}}{A_{\Phi}}\right)^2\ftNL I_{31}(M)$. 

Once we have these $I_{ij}$ values, we can now also study how much do the scaling of higher moments of smoothed moments deviate from the naive expectations, Eq.(\ref{eq:hierMn}) and Eq.(\ref{eq:feederMn}). For that, we look at the limit of small non-Gaussianity and therefore $\langle \delta_M^2\rangle \approx I_{21}(M)$. Then, for the single field hierarchical case, one simply gets: $\mathcal{M}_{n,M}^h \approx \fNL^{n-2} I_{n1}/I_{21}^{n/2}$, and for the feeder case, one gets: $\mathcal{M}_{n}^f \approx \left(q\ftNL\right)^{n}I_{n2}/I_{21}^{n/2}$. For the feeder case, we find the scaling of higher moments of the smoothed density field is only slightly different from the expectation Eq.(\ref{eq:feederMn}). We get $\mathcal{M}_{n,M}^f \approx  2^{n-1}(n-1)! \left(\frac{1.32 \mathcal{M}_3}{8}\right)^{n/3}$, for the range of mass scale ($\approx 10^{13}$ to $10^{15}$) $h^{-1}M_{\odot}$. In the same mass range, we find that the hierarchical case, similarly, satisfies a modified relation:  $\mathcal{M}_{n}^h \approx 2^{n-3}n! \left(\frac{1.58 \mathcal{M}_3}{6}\right)^{n-2}$, but the extra factor (here $1.58$ taken near $M=10^{14} h^{-1} M_{\odot}$) is weakly $M$ dependent (at a few percent level).

\section{Truncation and Error}
\label{app:erroranalysis}
For the hierarchical scaling, we will truncate the series to $N=s$ terms and call the result the $N=s$ truncation. This will produce a series with terms of order $\mathcal{M}_3^N$. For the feeder scaling, we will truncate the series at $N=s+2$ terms and call this the $N=s+2$ truncation; this will produce a series with terms of order $\mathcal{M}_{3}^{N/3}$. For both scalings, our definition of the $N$th term in the series follows the definition in \cite{Shandera2012}.

The utility of the Edgeworth series formalism lies in the fact that the PDF in Eq.(\ref{eq:pdf}) is an asymptotic series. Therefore, one can estimate the error induced by truncating the PDF to $N$th order by simply looking at the next term in the series. For the error analysis, we will adopt methods similar to that of \cite{Shandera2012} and look at the maximum $\mathcal{M}_3$ values (for both scalings) that the PDF can be computed with reasonable accuracy (20 percent) for the $\nu$ range that encompasses the halo masses that we will use from our simulation outputs. We will also look at the error as a function of $\nu$ for various $\mathcal{M}_3$ values relevant for our simulations. 

\begin{figure}
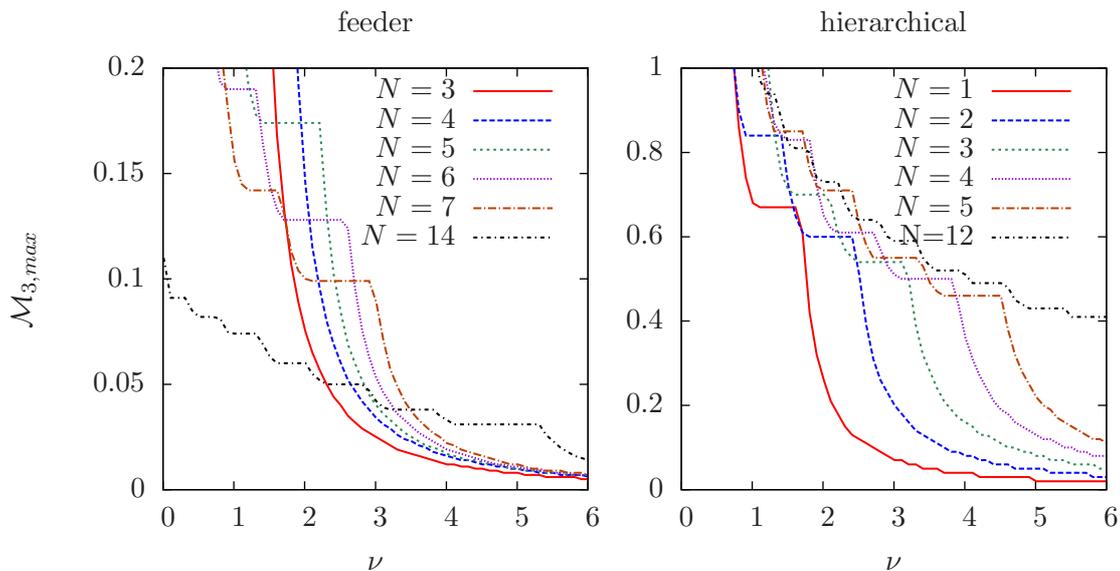

\centering 
\begin{subfigure}[c]{0.44\textwidth}
 \centering
 \input{figures/lm3maxf.tex}
\end{subfigure}
\begin{subfigure}[c]{0.55\textwidth}
 \centering
 \input{figures/lm3maxh.tex}
\end{subfigure}
\caption{Left: we plot the maximum value of $\mathcal{M}_3$ for which the PDF (for various truncations $N$) produces results within $20\%$ error for $\nu$ specified on the x-axis to $\nu_{max}=2.1\nu^{0.7}$ for feeder scaling of higher moments. Right: same as left but for hierarchical scaling and with $\nu_{max}=2.2\nu^{0.7}$.}
\label{trerror}
\end{figure}

\begin{figure}
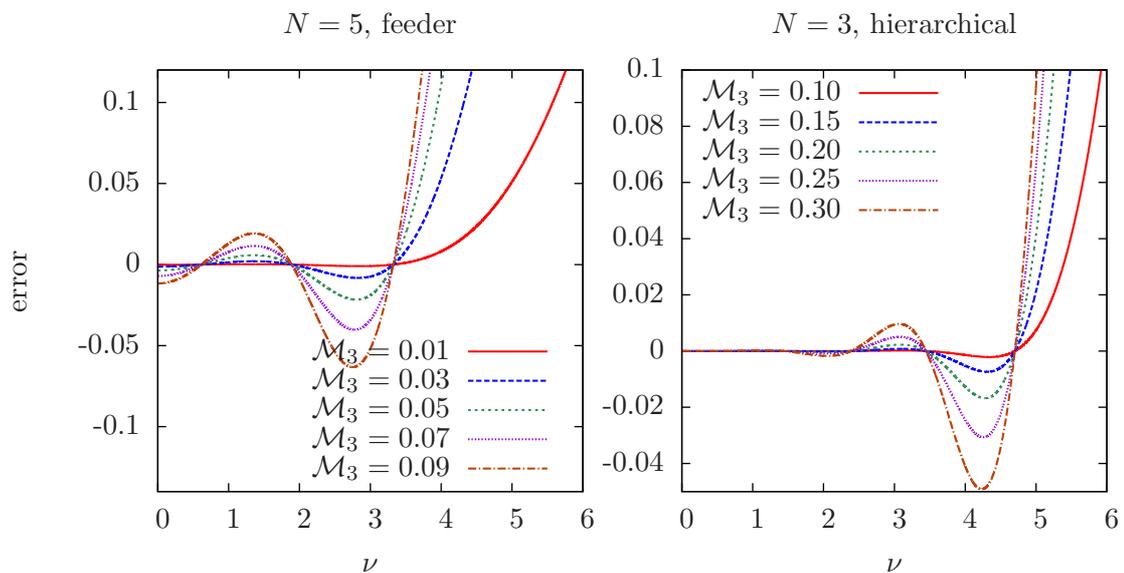

 \centering
 \begin{subfigure}[b]{0.45\textwidth}
  \input{figures/err3m3f.tex}
 \end{subfigure}
 \begin{subfigure}[b]{0.53\textwidth}
  \input{figures/err3m3h.tex}
 \end{subfigure}
 \caption{Left: the error of $N=5$ truncation for different values of $\mathcal{M}_3$ for feeder scaling. Right: same as left but for hierarchical scaling.}
 \label{trerrorM3}
\end{figure}

We find that the truncations of the PDF for the feeder scaling generate errors of magnitude $>20\%$ for much smaller $\nu \approx 3-4$ when keeping similar order terms in $\mathcal{M}_n$ compared to the hierarchical case (typically $\nu \approx 5-6$ for $n=5$). From Figure \ref{trerror}, we can also see that the value of $\mathcal{M}_{3,max}$ is much smaller for the feeder scaling and decreases sharply as one increases $\nu$. This means that at a higher mass range or at higher redshift, our simulation results may not be well described by our analytical formula for the non-Gaussian feeder mass function. Looking at the result for $N=14$, we see that one gains only marginally by increasing the number of terms for the feeder scaling. So, we will adopt $N=5$ truncation for the feeder case (i.e. up to $\mathcal{M}_3^{5/3}$) to compare with the simulation results. For the hierarchical scaling, we will adopt $N=3$ truncation (i.e up to $\mathcal{M}_3^5$). 

We can also see that the error increases as one increases $\nu$ or $\mathcal{M}_3$ (see Figure \ref{trerrorM3}). So, we expect the analytic mass function to describe simulation results better when the level of non-Gaussianity is smaller and at smaller $\nu_c$ i.e. low redshift and small halo masses.  

Also note that our error discussion assumes scalings: $\mathcal{M}_n^{\rm hier} = A_n (\mathcal{M}_3/6)^{n-2}$ and $\mathcal{M}_n^{\rm feeder} = B_n(\mathcal{M}_3/8)^{n/3}$. But from our calculated moments, we find that the scaling for the smoothed moments is modified slightly and the higher moments are larger than expected from the simple scaling assumed in the error analysis plots (see Appendix \ref{integrals}). The qualitative discussion remains the same but the magnitude of $\mathcal{M}_{3,max}$ will decrease and the relative error will increase in Figures \ref{trerror} and \ref{trerrorM3} respectively.

\section{Calculations for bias and stochasticity}
\label{app:bias}
Here we calculate the large scale bias and stochasticity expressions for our two field ansatz (\ref{twofieldmodel}). We take as a starting point the derivation for generic non-Gaussian scenarios given in \cite{Baumann2012}. The leading contribution to the matter-halo cross spectrum, in the long wavelength limit ($k\rightarrow0$) is
\begin{eqnarray}
P_{mh}(k)&=& b_\phi \left(\alpha^2(k) P_{\phi,G}(k) \right)+ \alpha^2(k) P_{\psi,\rm NG}(k)\left[b_\psi+\frac{1}{\alpha(k)}\left(\frac{1}{2}(b_\psi-1)\delta_c+\frac{1}{2}\frac{d}{d\ln\sigma_R}\right)\mathcal{F}_R^{(3)}\right] \nonumber \\ \label{eq:Pmh}
\\\nonumber
\mathcal{F}_R^{(3)}&=&\frac{1}{P_{\Phi}(k)\sigma_R^2}\int \frac{d^3 \vec{p_1}}{(2\pi)^3}\frac{d^3 \vec{p_2}}{(2\pi)^3}\,\alpha_R(p_1)\alpha_R(p_2)\langle\Phi(\vec{k})\Phi(\vec{p_1})\Phi(\vec{p_2})\rangle_c 
\end{eqnarray}
where $\alpha_R(k)=W_R(k)\alpha(k)$. This expression agrees with the result previously derived in \cite{Desjacques2011}. For our bispectrum, Eq.(\ref{eq:bispectrum}), there are two terms in $\mathcal{F}_R^{(3)}$. In the $k\rightarrow0$ limit, the usual term (proportional to $\ftNL$) is
\begin{eqnarray}
\mathcal{F}_{R,1}^{(3)}&=&\frac{4 \ftNL q^2}{[1+\ftNL^2 q \mathcal{P}_{\psi,G}(k) I_1(k)]}\frac{1}{\sigma_R^2}\int\frac{d^3 \vec{p}}{(2\pi)^3}\frac{\alpha_R(p)^2P_{\Phi}(p)}{[1+{\ftNL^2 q \mathcal{P}_{\psi,G}(p) I_1(p)}]} \label{eq:approx1}\\\nonumber
&\approx& 4 \ftNL q^2\\\nonumber
\end{eqnarray}
where the second line holds only if the non-Gaussian correction to the total power is negligible. In the above expression, we have used,
\begin{equation}
P_{\psi, G}(k)=\frac{q\,P_{\Phi}(k)}{1+\ftNL^2q\mathcal{P}_{\psi,G}I_1(k)}
\end{equation}
which relates the Gaussian power in the $\psi,G$ field to the total power, $P_{\Phi}$.

The second term, which is usually dropped as small in single field scenarios, is
\begin{eqnarray}
\mathcal{F}_{R,2}^{(3)}&=&\frac{8\ftNL^3}{P_{\Phi}(k)\sigma_R^2}\int \frac{d^3 \vec{p_1}}{(2\pi)^3} \alpha_R(p_1)\alpha_R(|\vec{p_1}+\vec{k}|) \int\frac{d^3 \vec{p}}{(2\pi)^3} P_{\psi, G}(p)P_{\psi, G}(|\vec{p}_1-\vec{p}|)P_{\psi, G}(|\vec{p}+\vec{k}|) \label{eq:FR32} \nonumber 
\end{eqnarray}

Let us now try to simplify the integral by looking at the major contributions to the integral. At large scales (small $k$), the value of $\mathcal{F}_R^{(3)}$ peaks when $p_1$ is near the halo scale ($\approx 1/R$). Typically, when looking at large scale bias, $k \ll p_1$. In this squeezed limit, we find that the loop bispectrum can be well approximated by:
\begin{eqnarray}
 \int \frac{d^3\vec{p}}{\tpc} P_{\psi, G}(p) P_{\psi,G}(|\vec{p_1}-\vec{p}|)P_{\psi,G}(|\vec{p}+\vec{k}|) \approx P_{\psi,G}(p_1) \int \frac{d^3\vec{p}}{\tpc} P_{\psi,G}(p) P_{\psi,G}(|\vec{p}+\vec{k}|) \nonumber \\
 \label{eq:approxbispectrum}
\end{eqnarray}
This is true because the dominant term for the integral comes from when $|\vec{p}+ \vec{k}|$ and $p\approx k$ are both small. This approximation breaks down as the ratio $p_1/k$ becomes smaller. However, we find that the dependence of the left hand side integral on the angular part of $\vec{p_1}$ is symmetric around the approximate value on the right hand side even when $p_1$ is only a few times larger than $k$. Since we integrate over $\vec{p_1}$ in the $\mathcal{F}_{R}^{(3)}$ integral, we expect the following approximation to hold quite well.
\begin{eqnarray}
 \mathcal{F}_{R,2}^{(3)}&\approx&\frac{8\ftNL^3}{P_{\Phi}(k)\sigma_R^2}\int \frac{d^3 \vec{p_1}}{(2\pi)^3} \alpha_R(p_1)^2 P_{\psi,G}(p_1) \int \frac{d^3 \vec{p}}{(2\pi)^3} P_{\psi, G}(p)P_{\psi, G}(|\vec{p}+\vec{k}|) \nonumber \\ 
 &=& \frac{8\ftNL^3}{P_{\Phi}(k)} \frac{ \sigma_{R,\psi}^2}{\sigma_R^2}  \int \frac{d^3 \vec{p}}{(2\pi)^3} P_{\psi, G}(p)P_{\psi, G}(|\vec{p}+\vec{k}|)
 \label{eq:approx2}
\end{eqnarray}

\begin{figure}
 \centering
 \input{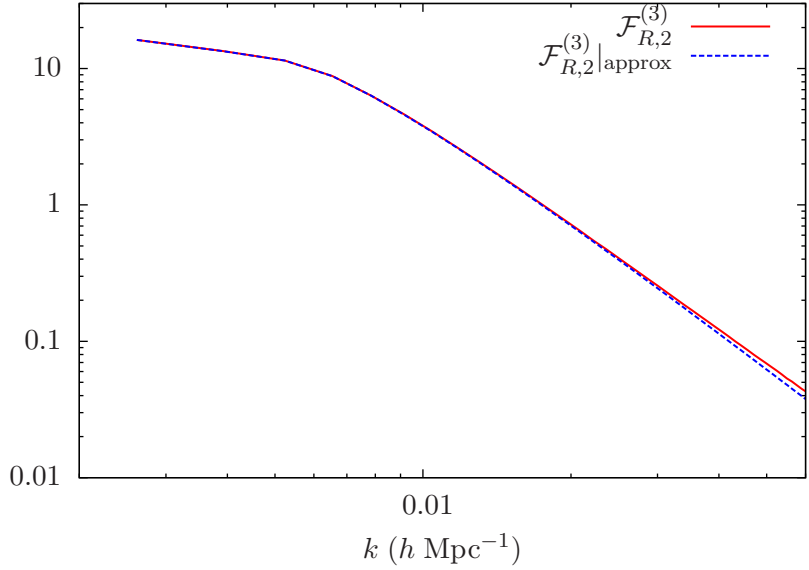}
 \caption{Here we test the approximation (\ref{eq:approx2}) for $\mathcal{F}_{R,2}^{(3)}$ numerically for $R=8\Mpch$. The approximation is excellent at large scales $k \lesssim 0.04 \hMpc$. Note that a different choice for $\ftNL$ and normalization for $P_{\psi,G}$ (or $q$) only rescales both curves by the same factor.}
 \label{fig:approx2}
\end{figure}

Approximation (\ref{eq:approx2}) was checked numerically using the CUBA library for multi-dimensional integration \cite{cuba}; the result is shown in Figure \ref{fig:approx2}. Further, we also tested that the derivative term in Eq.(\ref{eq:Pmh}) is indeed small compared to relevant values of $(b_g-1)\delta_c \mathcal{F}_{R}^{(3)}$. The derivative term for feeder scaling $\frac{d}{d \ln{\sigma_R}} \mathcal{F}_{R,2}^{(3)}$ was found to be of the same order as the derivative term for hierarchical scaling $\frac{d}{d \ln{\sigma_R}} \mathcal{F}_{R,1}^{(3)}$.

The approximated expression for bias can also be obtained directly from a peak background split analysis of our model. Following the peak background split derivation of \cite{Baumann2012} (section 4.1.1), as usual, we split both our Gaussian fields into short and long wavelength modes: $\phi_G(\vec{x}) = \phi_{G,l}(\vec{x}) + \phi_{G,s}(\vec{x})$ and $\psi_G(\vec{x})=\psi_{G,l}(\vec{x})+\psi_{G,s}(\vec{x})$. However, we also include the $\psi(\vec{x})^2$ term and the local small scale power is given by: $\sigma_R^2 = \bar{\sigma_{R}^2}\left[1+ 4 \frac{\sigma_{\psi,R}^2}{\sigma_{R}^2} \ftNL \psi_l(\vec{x}) \left(1 + \ftNL \psi_l(\vec{x}\right)\right]$. Then, allowing for separate linear bias coefficients for our two independent fields, we get:
\begin{eqnarray}
 \delta_h(\vec{k}) &=& b_\phi \delta_\phi (\vec{k}) + b_\psi \delta_{\psi,\rm NG} + 2 \delta_c (b_\psi-1) \ftNL \frac{\sigma_{R,\psi}^2}{\sigma_R^2} 
\left(\psi_l(\vec{k})+\ftNL \int \frac{d^3\vec{s}}{\tpc} \psi_l(\vec{s}) 
\psi_l(\vec{k}-\vec{s}) \right) \nonumber \\
&=& b_\phi \delta_\phi (\vec{k}) + \left(b_\psi + 2\delta_c (b_\psi-1) \frac{\ftNL}{\alpha(k)} \frac{\sigma_{R,\psi}^2}{\sigma_R^2}\right)\delta_{\psi,\rm NG}
\label{eq:deltah}
\end{eqnarray}
where 
\begin{eqnarray}
 \delta_{\phi}(\vec{k}) &=& \alpha(k) \phi_l(\vec{k})\\
 \delta_{\psi, \rm NG} &=& \alpha(k) \psi_l(\vec{k}) + \ftNL \int \frac{d^3\vec{p}}{\tpc} \alpha(k) \psi_l(\vec{p}) \psi_l(\vec{k}-\vec{p})
\end{eqnarray}
and therefore, the total linear matter density field is,
\begin{eqnarray}
 \delta_m(\vec{k})&=& \delta_{\phi}(\vec{k}) + \delta_{\psi, \rm NG}
\label{eq:deltam}
\end{eqnarray}

This gives the expression for $P_{hm}(k)$ to be,
\begin{eqnarray}
P_{hm}(k) &=& b_\phi \alpha^2(k)P_{\phi}(k) + \left(b_\psi + 2 \delta_c (b_\psi-1) \frac{\ftNL}{\alpha(k)} \frac{\sigma_{R,\psi}^2}{\sigma_R^2} \right) \alpha^2(k) P_{\psi, \rm NG}
\label{eq:hmbias}
 \end{eqnarray}
which agrees to the bias for our two field model given by Eq.(\ref{eq:Pmh}) after using the approximations for $\mathcal{F}_{R,1}^{(3)}$ and $\mathcal{F}_{R,2}^{(3)}$ from Eq.(\ref{eq:approx1}) and Eq.(\ref{eq:approx2}). Note that the factor $\sigma_{R,\psi}^2/\sigma_{R}^2 \approx q$ for the case of small non-Gaussianity; in this limit one gets the simpler expression Eq.(\ref{eq:biassimple}) but since we do not always stay in the limit of small non-Gaussianity in our simulations, we compute  this factor given a model specified by $q, \ftNL$ in our analysis.

Now to compute the stochastic bias, defined as
\begin{eqnarray}
 r^2(k) &=& \frac{P_{hm}^2(k)}{P_{hh}(k) P_{mm}(k)}
\end{eqnarray}
we need the halo-halo power spectrum. The leading contributions to the halo-halo power spectrum are
\begin{eqnarray}
P_{hh}(k)&=& b_\phi^2 (\alpha^2(k) P_{\phi,G}(k)) + \alpha^2(k) P_{\psi,NG}(k)\left[b_\psi^2+\frac{2 b_\psi}{\alpha(k)}\left(\frac{1}{2}(b_\psi-1)\delta_c+\frac{1}{2}\frac{d}{d\ln\sigma_R}\right)\mathcal{F}_R^{(3)}\right. \nonumber \\
&&\left.+\frac{1}{\alpha(k)^2}\left(\frac{1}{2}(b_\psi-1)\delta_c+\frac{1}{2}\frac{d}{d\ln\sigma_R}\right)^2\mathcal{F}_R^{(4)}\right]\\\nonumber
\mathcal{F}_R^{(4)}&=&\frac{1}{P_{\Phi}(k)\sigma_R^4}\int \frac{d^3\vec{p_1}}{(2\pi)^3}\frac{d^3 \vec{p_2}}{(2\pi)^3}\,\alpha_R^2(p_1)\alpha_R^2(p_2)\langle\Phi(\vec{p_1})\Phi(\vec{k}-\vec{p}_1)\Phi(\vec{p_2})\Phi(-\vec{k}-\vec{p}_2)\rangle_c
\end{eqnarray}
From our trispectrum, Eq.(\ref{eq:trispectrum}), there are two terms in $\mathcal{F}_R^{(4)}$. In the $k\rightarrow0$ limit, the usual term (proportional to $\ftNL^2$) is
\begin{eqnarray}
\mathcal{F}_{R,1}^{(4)}&=&\frac{16\ftNL^2 q^3}{[1+q\ftNL^2 I_1(k)\mathcal{P}_{\psi,G}(k)]}\left[\frac{1}{\sigma_R^2}\int\frac{d^3p}{(2\pi)^3}\frac{\alpha_R(p)^2P_{\Phi}(p)}{[1+{q\ftNL^2}I_1(p)\mathcal{P}_{\psi,G}(p)]}\right]^2\\\nonumber
&\approx& 16 \ftNL^2 q^3\\\nonumber
\end{eqnarray}

The other term, usually much smaller than the first term in single field cases, is

\begin{eqnarray}
 \mathcal{F}_{R,2}^{(4)} &=& \frac{48 \ftNL^4}{P_{\Phi}(k)\sigma_R^4} \int \frac{d^3\vec{p_1}}{\tpc} \frac{d^3 \vec{p_2}}{\tpc} \alpha_R^2(p_1) \alpha_R^2(p_2) \int \frac{d^3\vec{p}}{\tpc} \nonumber \\ & & \left[ P_{\psi,G}(p)P_{\psi,G}(|\vec{p_1}-\vec{p}|) P_{\psi, G}(|\vec{k}-\vec{p_1}+\vec{p}|) P_{\psi, G}(|\vec{p}-\vec{p_1}-\vec{p_2}|)  \right]
\end{eqnarray}

Similar to the case of $\mathcal{F}_{R,2}^{(3)}$, we can approximate this integral at large scales by looking at the collapsed limit: $k \ll p_1,p_2$. In this limit, the loop trispectrum is well approximated by:
\begin{eqnarray}
 \int \frac{d^3\vec{p}}{\tpc} P_{\psi,G}(p)P_{\psi,G}(|\vec{p_1}-\vec{p}|) P_{\psi, G}(|\vec{k}-\vec{p_1}+\vec{p}|) P_{\psi, G}(|\vec{p}-\vec{p_1}-\vec{p_2}|) &\approx& \nonumber \\  P_{\psi,G}(p_2) \int \frac{d^3\vec{p}}{\tpc} P_{\psi,G}(p) P_{\psi,G}(|\vec{p_1}-\vec{p}|)P_{\psi,G}(|-\vec{p_1}+\vec{k}+\vec{p}|)
\end{eqnarray}
following from the observation that the integral gets maximum contribution from when $|\vec{p_1}-\vec{p}|$ is small. Further, when $p_2/k$ is small, the approximation breaks down as before, but since we integrate over $\vec{p_2}$ to obtain $\mathcal{F}_{R,2}^{(4)}$, the symmetry of the integrand with respect to the angular part of $p_2$ will justify the use of this approximation in  $\mathcal{F}_{R,2}^{(4)}$ even when the halo formation scale is only a few times larger than the scale $k$ at which stochastic bias is measured.xx

With this approximation $\mathcal{F}_{R,2}^{(4)}$ becomes,
\begin{eqnarray}
 \mathcal{F}_{R,2}^{(4)} &\approx& \frac{48 \ftNL^4 \sigma_{\psi,R}^2}{P_{\Phi}(k)\sigma_R^4} \int \frac{d^3\vec{p_1}}{\tpc}  \alpha_R^2(p_1) \int \frac{d^3\vec{p}}{\tpc} P_{\psi,G}(p)P_{\psi,G}(|\vec{p_1}-\vec{p}|) P_{\psi, G}(|\vec{k}-\vec{p_1}+\vec{p}|)  \nonumber \\
\end{eqnarray}

These integrals are computationally challenging. However, as in the case of $P_{hm}(k)$, we expect the peak-background-split calculation to provide good approximation, in the large scale limit, to the halo-halo power spectrum. For this, using Eq.(\ref{eq:deltah}), one obtains,
\begin{eqnarray}
 P_{hh}(k) &=& b_\phi^2 (\alpha^2(k)P_{\phi}(k)) + \left(b_\psi+2\delta_c (b_\psi-1) \frac{\ftNL}{\alpha(k)}\frac{\sigma_{R,\psi}^2}{\sigma_R^2}\right)^2 \alpha^2(k) P_{\psi, \rm NG}
 \label{eq:hhbias}
\end{eqnarray}

To summarize this section, we have derived expressions Eq.(\ref{eq:hmbias}) (for $P_{hm}(k)$) and Eq.(\ref{eq:hhbias}) (for $P_{hh}(k)$), which were used to fit to our simulation results for large scale bias and stochastic bias (Eq.(\ref{eq:stochasticity})).

\bibliographystyle{JHEP}

\begin{thebibliography}{10}

\bibitem{PlanckCollaboration2013}
{\bf Planck} Collaboration, P.~Ade et~al., {\it {Planck 2013 Results. XXIV.
  Constraints on primordial non-Gaussianity}},
  \href{http://xxx.lanl.gov/abs/1303.5084}{{\tt arXiv:1303.5084}}.

\bibitem{Pillepich2012}
A.~Pillepich, C.~Porciani, and T.~H. Reiprich, {\it {The X-ray cluster survey
  with eROSITA: forecasts for cosmology, cluster physics, and primordial
  non-Gaussianity}},  {\em Mon.Not.Roy.Astron.Soc.} {\bf 422} (2012) 44--69,
  [\href{http://xxx.lanl.gov/abs/1111.6587}{{\tt arXiv:1111.6587}}].

\bibitem{Cunha2010}
C.~Cunha, D.~Huterer, and O.~Dore, {\it {Primordial non-Gaussianity from the
  covariance of galaxy cluster counts}},  {\em Phys.Rev.} {\bf D82} (2010)
  023004, [\href{http://xxx.lanl.gov/abs/1003.2416}{{\tt arXiv:1003.2416}}].

\bibitem{Oguri2009}
M.~Oguri, {\it {Self-Calibrated Cluster Counts as a Probe of Primordial
  Non-Gaussianity}},  {\em Phys.Rev.Lett.} {\bf 102} (2009) 211301,
  [\href{http://xxx.lanl.gov/abs/0905.0920}{{\tt arXiv:0905.0920}}].

\bibitem{Giannantonio2012}
T.~Giannantonio, C.~Porciani, J.~Carron, A.~Amara, and A.~Pillepich, {\it
  {Constraining primordial non-Gaussianity with future galaxy surveys}},  {\em
  Mon.Not.Roy.Astron.Soc.} {\bf 422} (2012) 2854--2877,
  [\href{http://xxx.lanl.gov/abs/1109.0958}{{\tt arXiv:1109.0958}}].

\bibitem{Merloni:2012uf}
A.~Merloni, P.~Predehl, W.~Becker, H.~Bohringer, T.~Boller, et~al., {\it
  {eROSITA Science Book: Mapping the Structure of the Energetic Universe}},
  \href{http://xxx.lanl.gov/abs/1209.3114}{{\tt arXiv:1209.3114}}.

\bibitem{Amendola:2012ys}
{\bf Euclid Theory Working Group} Collaboration, L.~Amendola et~al., {\it
  {Cosmology and fundamental physics with the Euclid satellite}},  {\em Living
  Rev.Rel.} {\bf 16} (2013) 6, [\href{http://xxx.lanl.gov/abs/1206.1225}{{\tt
  arXiv:1206.1225}}].

\bibitem{Jeong:2009vd}
D.~Jeong and E.~Komatsu, {\it {Primordial non-Gaussianity, scale-dependent
  bias, and the bispectrum of galaxies}},  {\em Astrophys.J.} {\bf 703} (2009)
  1230--1248, [\href{http://xxx.lanl.gov/abs/0904.0497}{{\tt
  arXiv:0904.0497}}].

\bibitem{Baldauf:2010vn}
T.~Baldauf, U.~Seljak, and L.~Senatore, {\it {Primordial non-Gaussianity in the
  Bispectrum of the Halo Density Field}},  {\em JCAP} {\bf 1104} (2011) 006,
  [\href{http://xxx.lanl.gov/abs/1011.1513}{{\tt arXiv:1011.1513}}].

\bibitem{Tasinato:2013vna}
G.~Tasinato, M.~Tellarini, A.~J. Ross, and D.~Wands, {\it {Primordial
  non-Gaussianity in the bispectra of large-scale structure}},
  \href{http://xxx.lanl.gov/abs/1310.7482}{{\tt arXiv:1310.7482}}.

\bibitem{Barnaby2012}
N.~Barnaby and S.~Shandera, {\it {Feeding your Inflaton: Non-Gaussian
  Signatures of Interaction Structure}},  {\em JCAP} {\bf 1201} (2012) 034,
  [\href{http://xxx.lanl.gov/abs/1109.2985}{{\tt arXiv:1109.2985}}].

\bibitem{Loverde2007}
M.~LoVerde, A.~Miller, S.~Shandera, and L.~Verde, {\it {Effects of
  Scale-Dependent Non-Gaussianity on Cosmological Structures}},  {\em JCAP}
  {\bf 0804} (2008) 014, [\href{http://xxx.lanl.gov/abs/0711.4126}{{\tt
  arXiv:0711.4126}}].

\bibitem{Shandera2013}
S.~Shandera, A.~Mantz, D.~Rapetti, and S.~W. Allen, {\it {X-ray Cluster
  Constraints on Non-Gaussianity}},  {\em JCAP} {\bf 1308} (2013) 004,
  [\href{http://xxx.lanl.gov/abs/1304.1216}{{\tt arXiv:1304.1216}}].

\bibitem{Trumper1993}
J.~{Truemper}, {\it {ROSAT - A new look at the X-ray sky}},  {\em Science} {\bf
  260} (June, 1993) 1769--1771.

\bibitem{Benson2013}
B.~Benson, T.~de~Haan, J.~Dudley, C.~Reichardt, K.~Aird, et~al., {\it
  {Cosmological Constraints from Sunyaev-Zel'dovich-Selected Clusters with
  X-ray Observations in the First 178 Square Degrees of the South Pole
  Telescope Survey}},  {\em Astrophys.J.} {\bf 763} (2013) 147,
  [\href{http://xxx.lanl.gov/abs/1112.5435}{{\tt arXiv:1112.5435}}].

\bibitem{Williamson2011}
R.~Williamson, B.~Benson, F.~High, K.~Vanderlinde, P.~Ade, et~al., {\it {An
  SZ-selected sample of the most massive galaxy clusters in the
  2500-square-degree South Pole Telescope survey}},  {\em Astrophys.J.} {\bf
  738} (2011) 139, [\href{http://xxx.lanl.gov/abs/1101.1290}{{\tt
  arXiv:1101.1290}}].

\bibitem{Mana:2013qba}
A.~Mana, T.~Giannantonio, J.~Weller, B.~Hoyle, G.~Huetsi, et~al., {\it
  {Combining clustering and abundances of galaxy clusters to test cosmology and
  primordial non-Gaussianity}},  \href{http://xxx.lanl.gov/abs/1303.0287}{{\tt
  arXiv:1303.0287}}.

\bibitem{Afshordi:2008ru}
N.~Afshordi and A.~J. Tolley, {\it {Primordial non-gaussianity, statistics of
  collapsed objects, and the Integrated Sachs-Wolfe effect}},  {\em Phys.Rev.}
  {\bf D78} (2008) 123507, [\href{http://xxx.lanl.gov/abs/0806.1046}{{\tt
  arXiv:0806.1046}}].

\bibitem{Slosar2010}
A.~Slosar, C.~Hirata, U.~Seljak, S.~Ho, and N.~Padmanabhan, {\it {Constraints
  on local primordial non-Gaussianity from large scale structure}},  {\em JCAP}
  {\bf 0808} (2008) 031, [\href{http://xxx.lanl.gov/abs/0805.3580}{{\tt
  arXiv:0805.3580}}].

\bibitem{Xia:2010pe}
J.-Q. Xia, A.~Bonaldi, C.~Baccigalupi, G.~De~Zotti, S.~Matarrese, et~al., {\it
  {Constraining Primordial Non-Gaussianity with High-Redshift Probes}},  {\em
  JCAP} {\bf 1008} (2010) 013, [\href{http://xxx.lanl.gov/abs/1007.1969}{{\tt
  arXiv:1007.1969}}].

\bibitem{Xia:2011hj}
J.-Q. Xia, C.~Baccigalupi, S.~Matarrese, L.~Verde, and M.~Viel, {\it
  {Constraints on Primordial Non-Gaussianity from Large Scale Structure
  Probes}},  {\em JCAP} {\bf 1108} (2011) 033,
  [\href{http://xxx.lanl.gov/abs/1104.5015}{{\tt arXiv:1104.5015}}].

\bibitem{Ross:2012sx}
A.~J. Ross, W.~J. Percival, A.~Carnero, G.-b. Zhao, M.~Manera, et~al., {\it
  {The Clustering of Galaxies in SDSS-III DR9 Baryon Oscillation Spectroscopic
  Survey: Constraints on Primordial Non-Gaussianity}},  {\em
  Mon.Not.Roy.Astron.Soc.} {\bf 428} (2013) 1116--1127,
  [\href{http://xxx.lanl.gov/abs/1208.1491}{{\tt arXiv:1208.1491}}].

\bibitem{Karagiannis:2013xea}
D.~Karagiannis, T.~Shanks, and N.~P. Ross, {\it {Search for primordial
  non-Gaussianity in the quasars of SDSS-III BOSS DR9}},
  \href{http://xxx.lanl.gov/abs/1310.6716}{{\tt arXiv:1310.6716}}.

\bibitem{Giannantonio:2013uqa}
T.~Giannantonio, A.~J. Ross, W.~J. Percival, R.~Crittenden, D.~Bacher, et~al.,
  {\it {Improved Primordial Non-Gaussianity Constraints from Measurements of
  Galaxy Clustering and the Integrated Sachs-Wolfe Effect}},
  \href{http://xxx.lanl.gov/abs/1303.1349}{{\tt arXiv:1303.1349}}.

\bibitem{Ho:2013lda}
S.~Ho, N.~Agarwal, A.~D. Myers, R.~Lyons, A.~Disbrow, et~al., {\it {Sloan
  Digital Sky Survey III Photometric Quasar Clustering: Probing the Initial
  Conditions of the Universe using the Largest Volume}},
  \href{http://xxx.lanl.gov/abs/1311.2597}{{\tt arXiv:1311.2597}}.

\bibitem{Agarwal:2013qta}
N.~Agarwal, S.~Ho, and S.~Shandera, {\it {Constraining the initial conditions
  of the Universe using large scale structure}},
  \href{http://xxx.lanl.gov/abs/1311.2606}{{\tt arXiv:1311.2606}}.

\bibitem{Pillepich:2008ka}
A.~Pillepich, C.~Porciani, and O.~Hahn, {\it {Universal halo mass function and
  scale-dependent bias from N-body simulations with non-Gaussian initial
  conditions}},  \href{http://xxx.lanl.gov/abs/0811.4176}{{\tt
  arXiv:0811.4176}}.

\bibitem{Grossi:2009an}
M.~Grossi, L.~Verde, C.~Carbone, K.~Dolag, E.~Branchini, et~al., {\it
  {Large-scale non-Gaussian mass function and halo bias: tests on N-body
  simulations}},  {\em Mon.Not.Roy.Astron.Soc.} {\bf 398} (2009) 321--332,
  [\href{http://xxx.lanl.gov/abs/0902.2013}{{\tt arXiv:0902.2013}}].

\bibitem{Giannantonio:2009ak}
T.~Giannantonio and C.~Porciani, {\it {Structure formation from non-Gaussian
  initial conditions: multivariate biasing, statistics, and comparison with
  N-body simulations}},  {\em Phys.Rev.} {\bf D81} (2010) 063530,
  [\href{http://xxx.lanl.gov/abs/0911.0017}{{\tt arXiv:0911.0017}}].

\bibitem{Loverde2011}
M.~LoVerde and K.~M. Smith, {\it {The Non-Gaussian Halo Mass Function with
  $f_{NL}$, $g_{NL}$ and $\tau_{NL}$}},  {\em JCAP} {\bf 1108} (2011) 003,
  [\href{http://xxx.lanl.gov/abs/1102.1439}{{\tt arXiv:1102.1439}}].

\bibitem{D'Amico:2010ta}
G.~D'Amico, M.~Musso, J.~Norena, and A.~Paranjape, {\it {An Improved
  Calculation of the Non-Gaussian Halo Mass Function}},  {\em JCAP} {\bf 1102}
  (2011) 001, [\href{http://xxx.lanl.gov/abs/1005.1203}{{\tt
  arXiv:1005.1203}}].

\bibitem{Wagner2010}
C.~Wagner, L.~Verde, and L.~Boubekeur, {\it {N-body simulations with generic
  non-Gaussian initial conditions I: Power Spectrum and halo mass function}},
  {\em JCAP} {\bf 1010} (2010) 022,
  [\href{http://xxx.lanl.gov/abs/1006.5793}{{\tt arXiv:1006.5793}}].

\bibitem{Dalal2008}
N.~Dalal, O.~Dore, D.~Huterer, and A.~Shirokov, {\it {The imprints of
  primordial non-gaussianities on large-scale structure: scale dependent bias
  and abundance of virialized objects}},  {\em Phys.Rev.} {\bf D77} (2008)
  123514, [\href{http://xxx.lanl.gov/abs/0710.4560}{{\tt arXiv:0710.4560}}].

\bibitem{Tseliakhovich2010}
D.~Tseliakhovich, C.~Hirata, and A.~Slosar, {\it {Non-Gaussianity and
  large-scale structure in a two-field inflationary model}},  {\em Phys.Rev.}
  {\bf D82} (2010) 043531, [\href{http://xxx.lanl.gov/abs/1004.3302}{{\tt
  arXiv:1004.3302}}].

\bibitem{Linde:1996gt}
A.~D. Linde and V.~F. Mukhanov, {\it {Nongaussian isocurvature perturbations
  from inflation}},  {\em Phys.Rev.} {\bf D56} (1997) 535--539,
  [\href{http://xxx.lanl.gov/abs/astro-ph/9610219}{{\tt astro-ph/9610219}}].

\bibitem{Moroi:2001ct}
T.~Moroi and T.~Takahashi, {\it {Effects of cosmological moduli fields on
  cosmic microwave background}},  {\em Phys.Lett.} {\bf B522} (2001) 215--221,
  [\href{http://xxx.lanl.gov/abs/hep-ph/0110096}{{\tt hep-ph/0110096}}].

\bibitem{Lyth2001}
D.~H. Lyth and D.~Wands, {\it {Generating the curvature perturbation without an
  inflaton}},  {\em Phys.Lett.} {\bf B524} (2002) 5--14,
  [\href{http://xxx.lanl.gov/abs/hep-ph/0110002}{{\tt hep-ph/0110002}}].

\bibitem{Enqvist:2001zp}
K.~Enqvist and M.~S. Sloth, {\it {Adiabatic CMB perturbations in pre - big bang
  string cosmology}},  {\em Nucl.Phys.} {\bf B626} (2002) 395--409,
  [\href{http://xxx.lanl.gov/abs/hep-ph/0109214}{{\tt hep-ph/0109214}}].

\bibitem{Dvali:2003em}
G.~Dvali, A.~Gruzinov, and M.~Zaldarriaga, {\it {A new mechanism for generating
  density perturbations from inflation}},  {\em Phys.Rev.} {\bf D69} (2004)
  023505, [\href{http://xxx.lanl.gov/abs/astro-ph/0303591}{{\tt
  astro-ph/0303591}}].

\bibitem{Zaldarriaga:2003my}
M.~Zaldarriaga, {\it {Non-Gaussianities in models with a varying inflaton decay
  rate}},  {\em Phys.Rev.} {\bf D69} (2004) 043508,
  [\href{http://xxx.lanl.gov/abs/astro-ph/0306006}{{\tt astro-ph/0306006}}].

\bibitem{Creminelli:2004yq}
P.~Creminelli and M.~Zaldarriaga, {\it {Single field consistency relation for
  the 3-point function}},  {\em JCAP} {\bf 0410} (2004) 006,
  [\href{http://xxx.lanl.gov/abs/astro-ph/0407059}{{\tt astro-ph/0407059}}].

\bibitem{Pajer2013}
E.~Pajer, F.~Schmidt, and M.~Zaldarriaga, {\it {The Observed Squeezed Limit of
  Cosmological Three-Point Functions}},  {\em Phys.Rev.} {\bf D88} (2013),
  no.~8 083502, [\href{http://xxx.lanl.gov/abs/1305.0824}{{\tt
  arXiv:1305.0824}}].

\bibitem{Keisler:2011aw}
R.~Keisler, C.~Reichardt, K.~Aird, B.~Benson, L.~Bleem, et~al., {\it {A
  Measurement of the Damping Tail of the Cosmic Microwave Background Power
  Spectrum with the South Pole Telescope}},  {\em Astrophys.J.} {\bf 743}
  (2011) 28, [\href{http://xxx.lanl.gov/abs/1105.3182}{{\tt arXiv:1105.3182}}].

\bibitem{Sievers:2013ica}
{\bf Atacama Cosmology Telescope} Collaboration, J.~L. Sievers et~al., {\it
  {The Atacama Cosmology Telescope: Cosmological parameters from three seasons
  of data}},  {\em JCAP} {\bf 1310} (2013) 060,
  [\href{http://xxx.lanl.gov/abs/1301.0824}{{\tt arXiv:1301.0824}}].

\bibitem{Ade:2013zuv}
{\bf Planck} Collaboration, P.~Ade et~al., {\it {Planck 2013 results. XVI.
  Cosmological parameters}},  \href{http://xxx.lanl.gov/abs/1303.5076}{{\tt
  arXiv:1303.5076}}.

\bibitem{Lyth2006}
D.~H. Lyth, {\it {Non-gaussianity and cosmic uncertainty in curvaton-type
  models}},  {\em JCAP} {\bf 0606} (2006) 015,
  [\href{http://xxx.lanl.gov/abs/astro-ph/0602285}{{\tt astro-ph/0602285}}].

\bibitem{Lyth2007}
D.~H. Lyth, {\it {The curvature perturbation in a box}},  {\em JCAP} {\bf 0712}
  (2007) 016, [\href{http://xxx.lanl.gov/abs/0707.0361}{{\tt
  arXiv:0707.0361}}].

\bibitem{Chen2010}
X.~Chen and Y.~Wang, {\it {Quasi-Single Field Inflation and
  Non-Gaussianities}},  {\em JCAP} {\bf 1004} (2010) 027,
  [\href{http://xxx.lanl.gov/abs/0911.3380}{{\tt arXiv:0911.3380}}].

\bibitem{Nelson2013}
E.~Nelson and S.~Shandera, {\it {Statistical Naturalness and non-Gaussianity in
  a Finite Universe}},  {\em Phys.Rev.Lett.} {\bf 110} (2013), no.~13 131301,
  [\href{http://xxx.lanl.gov/abs/1212.4550}{{\tt arXiv:1212.4550}}].

\bibitem{Loverde2013}
M.~LoVerde, E.~Nelson, and S.~Shandera, {\it {Non-Gaussian Mode Coupling and
  the Statistical Cosmological Principle}},  {\em JCAP} {\bf 1306} (2013) 024,
  [\href{http://xxx.lanl.gov/abs/1303.3549}{{\tt arXiv:1303.3549}}].

\bibitem{Press1974}
W.~H. Press and P.~Schechter, {\it {Formation of galaxies and clusters of
  galaxies by selfsimilar gravitational condensation}},  {\em Astrophys.J.}
  {\bf 187} (1974) 425--438.

\bibitem{Tinker2008}
J.~L. Tinker, A.~V. Kravtsov, A.~Klypin, K.~Abazajian, M.~S. Warren, et~al.,
  {\it {Toward a halo mass function for precision cosmology: The Limits of
  universality}},  {\em Astrophys.J.} {\bf 688} (2008) 709--728,
  [\href{http://xxx.lanl.gov/abs/0803.2706}{{\tt arXiv:0803.2706}}].

\bibitem{petrovpaper}
V.~V. Petrov, {\em Sums of independent random variables.}
\newblock Springer-Verlag, 1975.

\bibitem{Blinnikov:1997jq}
S.~Blinnikov and R.~Moessner, {\it {Expansions for nearly Gaussian
  distributions}},  {\em Astron.Astrophys.Suppl.Ser.} {\bf 130} (1998)
  193--205, [\href{http://xxx.lanl.gov/abs/astro-ph/9711239}{{\tt
  astro-ph/9711239}}].

\bibitem{Bernardeau:2001qr}
F.~Bernardeau, S.~Colombi, E.~Gaztanaga, and R.~Scoccimarro, {\it Large-scale
  structure of the universe and cosmological perturbation theory},  {\em Phys.
  Rept.} {\bf 367} (2002) 1--248,
  [\href{http://xxx.lanl.gov/abs/astro-ph/0112551}{{\tt astro-ph/0112551}}].

\bibitem{Matarrese:1986et}
S.~Matarrese, F.~Lucchin, and S.~A. Bonometto, {\it {A Path Integral Approach
  To Large Scale Matter Distribution Originated By Nongaussian Fluctuations}},
  {\em Astrophys.J.} {\bf 310} (1986) L21--L26.

\bibitem{Matarrese:2008nc}
S.~Matarrese and L.~Verde, {\it {The effect of primordial non-Gaussianity on
  halo bias}},  {\em Astrophys.J.} {\bf 677} (2008) L77--L80,
  [\href{http://xxx.lanl.gov/abs/0801.4826}{{\tt arXiv:0801.4826}}].

\bibitem{Desjacques2011}
V.~Desjacques, D.~Jeong, and F.~Schmidt, {\it {Accurate Predictions for the
  Scale-Dependent Galaxy Bias from Primordial Non-Gaussianity}},  {\em
  Phys.Rev.} {\bf D84} (2011) 061301,
  [\href{http://xxx.lanl.gov/abs/1105.3476}{{\tt arXiv:1105.3476}}].

\bibitem{Desjacques:2011mq}
V.~Desjacques, D.~Jeong, and F.~Schmidt, {\it {Non-Gaussian Halo Bias
  Re-examined: Mass-dependent Amplitude from the Peak-Background Split and
  Thresholding}},  {\em Phys.Rev.} {\bf D84} (2011) 063512,
  [\href{http://xxx.lanl.gov/abs/1105.3628}{{\tt arXiv:1105.3628}}].

\bibitem{Baumann2012}
D.~Baumann, S.~Ferraro, D.~Green, and K.~M. Smith, {\it {Stochastic Bias from
  Non-Gaussian Initial Conditions}},  {\em JCAP} {\bf 1305} (2013) 001,
  [\href{http://xxx.lanl.gov/abs/1209.2173}{{\tt arXiv:1209.2173}}].

\bibitem{Yokoyama2011}
S.~Yokoyama, {\it {Scale-dependent bias from the primordial non-Gaussianity
  with a Gaussian-squared field}},  {\em JCAP} {\bf 1111} (2011) 001,
  [\href{http://xxx.lanl.gov/abs/1108.5569}{{\tt arXiv:1108.5569}}].

\bibitem{Smith2011}
K.~M. Smith and M.~LoVerde, {\it {Local stochastic non-Gaussianity and N-body
  simulations}},  {\em JCAP} {\bf 1111} (2011) 009,
  [\href{http://xxx.lanl.gov/abs/1010.0055}{{\tt arXiv:1010.0055}}].

\bibitem{gadget2}
V.~Springel, {\it {The Cosmological simulation code GADGET-2}},  {\em
  Mon.Not.Roy.Astron.Soc.} {\bf 364} (2005) 1105--1134,
  [\href{http://xxx.lanl.gov/abs/astro-ph/0505010}{{\tt astro-ph/0505010}}].

\bibitem{Crocce:2006ve}
M.~Crocce, S.~Pueblas, and R.~Scoccimarro, {\it {Transients from Initial
  Conditions in Cosmological Simulations}},  {\em Mon.Not.Roy.Astron.Soc.} {\bf
  373} (2006) 369--381, [\href{http://xxx.lanl.gov/abs/astro-ph/0606505}{{\tt
  astro-ph/0606505}}].

\bibitem{ahf}
S.~R. Knollmann and A.~Knebe, {\it {Ahf: Amiga's Halo Finder}},  {\em
  Astrophys.J.Suppl.} {\bf 182} (2009) 608--624,
  [\href{http://xxx.lanl.gov/abs/0904.3662}{{\tt arXiv:0904.3662}}].

\bibitem{Maggiore:2009rx}
M.~Maggiore and A.~Riotto, {\it {The Halo mass function from excursion set
  theory. III. Non-Gaussian fluctuations}},  {\em Astrophys.J.} {\bf 717}
  (2010) 526--541, [\href{http://xxx.lanl.gov/abs/0903.1251}{{\tt
  arXiv:0903.1251}}].

\bibitem{Maggiore:2009hp}
M.~Maggiore and A.~Riotto, {\it {The Halo Mass Function from Excursion Set
  Theory with a Non-Gaussian Trispectrum}},  {\em Mon.Not.Roy.Astron.Soc.Lett.}
  {\bf 405} (2010) 1244--1252, [\href{http://xxx.lanl.gov/abs/0910.5125}{{\tt
  arXiv:0910.5125}}].

\bibitem{DeSimone:2010mu}
A.~De~Simone, M.~Maggiore, and A.~Riotto, {\it {Excursion Set Theory for
  generic moving barriers and non-Gaussian initial conditions}},  {\em
  Mon.Not.Roy.Astron.Soc.} {\bf 412} (2011) 2587,
  [\href{http://xxx.lanl.gov/abs/1007.1903}{{\tt arXiv:1007.1903}}].

\bibitem{DeSimone:2011dn}
A.~De~Simone, M.~Maggiore, and A.~Riotto, {\it {Conditional Probabilities in
  the Excursion Set Theory. Generic Barriers and non-Gaussian Initial
  Conditions}},  {\em Mon.Not.Roy.Astron.Soc.} {\bf 418} (2011) 2403,
  [\href{http://xxx.lanl.gov/abs/1102.0046}{{\tt arXiv:1102.0046}}].

\bibitem{Musso:2013pja}
M.~Musso and R.~K. Sheth, {\it {The excursion set approach in non-Gaussian
  random fields}},  \href{http://xxx.lanl.gov/abs/1305.0724}{{\tt
  arXiv:1305.0724}}.

\bibitem{Achitouv:2013oea}
I.~Achitouv, C.~Wagner, J.~Weller, and Y.~Rasera, {\it {Computation of the Halo
  Mass Function Using Physical Collapse Parameters: Application to Non-Standard
  Cosmologies}},  \href{http://xxx.lanl.gov/abs/1312.1364}{{\tt
  arXiv:1312.1364}}.

\bibitem{Baldauf2013}
T.~Baldauf, U.~Seljak, R.~E. Smith, N.~Hamaus, and V.~Desjacques, {\it {Halo
  stochasticity from exclusion and nonlinear clustering}},  {\em Phys.Rev.}
  {\bf D88} (2013), no.~8 083507,
  [\href{http://xxx.lanl.gov/abs/1305.2917}{{\tt arXiv:1305.2917}}].

\bibitem{Desjacques2009}
V.~Desjacques, U.~Seljak, and I.~Iliev, {\it {Scale-dependent bias induced by
  local non-Gaussianity: A comparison to N-body simulations}},  {\em
  Mon.Not.Roy.Astron.Soc.} {\bf 396} (2009) 85--96,
  [\href{http://xxx.lanl.gov/abs/0811.2748}{{\tt arXiv:0811.2748}}].

\bibitem{Shandera2012}
S.~Shandera, A.~L. Erickcek, P.~Scott, and J.~Y. Galarza, {\it {Number Counts
  and Non-Gaussianity}},  {\em Phys.Rev.} {\bf D88} (2013), no.~10 103506,
  [\href{http://xxx.lanl.gov/abs/1211.7361}{{\tt arXiv:1211.7361}}].

\bibitem{cuba}
T.~Hahn, {\it {CUBA: A Library for multidimensional numerical integration}},
  {\em Comput.Phys.Commun.} {\bf 168} (2005) 78--95,
  [\href{http://xxx.lanl.gov/abs/hep-ph/0404043}{{\tt hep-ph/0404043}}].

\end{thebibliography}
\begingroup\raggedright\endgroup
\end{document}